\documentclass[apj]{emulateapj}
\usepackage{apjfonts}
\usepackage{lscape}

\begin{document}


\title{Bulges Of Nearby Galaxies With Spitzer: The Growth Of
  Pseudobulges In Disk Galaxies And Its Connection To Outer Disks}

\shorttitle{Star Formation in Bulges}
\shortauthors{Fisher, Drory \& Fabricius}


\author{David~B.~Fisher\altaffilmark{1}, Niv~Drory\altaffilmark{2} \&
  Maximilian H. Fabricius\altaffilmark{2} \altaffilmark{3}}

\email{dbfisher@astro.as.utexas.edu}

\altaffiltext{1}{Department of Astronomy, The University of Texas at Austin,
  1 University Station C1400, Austin, Texas 78712}

\altaffiltext{2}{Max-Planck-Institut f\"ur
  Extraterrestrische Physik, Giessenbachstra\ss e, Garching, Germany}

\altaffiltext{3}{University Observatory Munich, Schienerstrasse 1,
  81679, Munich, Germany}

\slugcomment{Submitted to ApJ}


\begin{abstract}
  We study star formation rates (SFR) and stellar masses in bulges of
  nearby disk galaxies. For this we construct a new SFR indicator that
  linearly combines data from Spitzer Space Telescope (SST) and The
  Galaxy Evolution Explorer (GALEX).  All bulges are found to be
  forming stars irrespective of bulge type (pseudobulge or classical
  bulge). At present day SFR the median pseudobulge could have grown
  the present day stellar mass in 8~Gyr. Classical bulges have the
  lowest specific SFR implying a growth times that are longer than a
  Hubble time, and thus the present day SFR does not likely play a
  major role in the evolution of classical bulges.  In almost all
  galaxies in our sample the specific SFR (SFR per unit stellar mass)
  of the bulge is higher than that of the outer disk. This suggests
  that almost all galaxies are increasing their $B/T$ through internal
  star formation.  SFR in pseudobulges correlates with their
  structure.  More massive pseudobulges have higher SFR density, this
  is consistent with that stellar mass being formed by moderate,
  extended star fromation.  Bulges in late-type galaxies have similar
  SFRs as pseudobulges in intermediate-type galaxies, and are similar
  in radial size. However, they are deficient in mass; thus, they have
  much shorter growth times, $\sim$2~Gyr. We identify a class of
  bulges that have nuclear morphology similar to pseudobulges,
  significantly lower specific SFR than pseudobulges, and are closer
  to classical bulges in structural parameter correlations. These are
  possibly composite objects, evolved pseudobulges or classical bulges
  experiencing transient, enhanced nuclear star formation. 

  Our results are consistent with a scenario in which bulge growth
  via internal star formation is a natural, and near ubiquitous
  phenomenon in disk galaxies. Those galaxies with large classical
  bulges are not affected by the {\em in situ} bulge growth, likely
  because the majority of their stellar mass comes from some other
  phenomenon. Yet, those galaxies with out a classical bulge, over
  long periods of extended star formation are able to growth a
  pseudobulge. Though cold accretion is not ruled out, for pseudobulge
  galaxies an addition of stellar mass from mergers or accretion is
  not required to explain the bulge mass. In this sense, galaxies with
  pseudobulges may very well be bulgeless (or ``quasi-bulgeless'')
  galaxies, and galaxies with classical bulges are galaxies in which
  both internal evolution and hierarchical merging are responsible for
  the bulge mass by fractions that vary from galaxy-to-galaxy.

\end{abstract}



\section{Introduction}\label{sec:introduction}

Observations now indicate that many bulges in nearby disk galaxies are
more complicated than previously thought. Bulges were once considered
to be essentially elliptical galaxies surrounded by disks
\citep[e.g.][]{renzini99}. Yet, contrary to historic assumptions
\citep[e.g.][]{whitford1978}, we now know that bulges are not
typically uniformly old, non-star forming systems.  Many bulges in the
nearby universe are filled with young stars and cold molecular gas
\citep{peletier1996,helfer2003}. The most active bulges have star
formation rates as high as 1~M$_{\odot}$~yr$^{-1}$
\citep{kk04,kennicutt1998b}, and exceedingly higher star formation
rate densities than their outer disk. Yet it is certainly true that
many non-star forming, red bulges exists; for example the nearest
bulge outside our own galaxy, M~31, is made of mostly old stars.
\cite{fisher2006} shows with mid-IR colors, bulges are either actively
forming stars at similar rates to their associated outer disk, or they
show a break in mid-IR colors indicating no activity in the bulge;
star formation activity in bulges is bimodal. \cite{fisher2006} finds
that those bulges with active disk-like star-formation have disk-like
morphology within the central few hundred parsecs of the bulge. Such
differences suggest that the nature of bulge growth in nearby disks
galaxies is dichotomous.  In this paper we wish to study the nature of
the mass growth in nearby star-forming bulges.  We compare present day
star formation rates to stellar masses of bulges to estimate the
significance of the star formation rate in nearby active bulges. We
also compare star formation rate density to properties of the bulge
and the disk.

In addition to stellar populations and SFR, bulges lack homogeneity in
many fundamental properties (see \citealp{kk04} for a review).  The
observations suggest that bulges are bimodal in nature, and
furthermore, this division is linked to the well-known bimodality in
global galaxy properties \citep{droryfisher2007}. Thus, it seems that
the difference in bulge type is fundamentally connected to the history
of the entire galaxy.

The two types of bulges are typically called {\em classical bulges}
and {\em pseudobulges}.  Classical bulges have properties similar to
elliptical galaxies; whereas pseudobulges are similar in many ways to
disk galaxies.  Properties that can identify bulges as pseudobulges
include the following: dynamics that are dominated by rotation
\citep{k93}, the bulge has a nearly exponential surface brightness
profile \citep{fisherdrory2008,spitzer1}, flattening similar to that
of their outer disk \citep{fathi2003,k93}, nuclear bar
\citep{erwin2002}, nuclear ring, and/or nuclear spiral
\citep{carollo97}. Classical bulges are typically identified as having
hot stellar dynamics, more nearly r$^{1/4}$ surface brightness
profiles, typically more round than the outer disk
\citep{fisherdrory2008}, and a relatively featureless morphology.
Also, \cite{fisherdrory2008} show that the structural properties of
classical bulges (absolute magnitude, S\'ersic index, half-light
radius, and surface brightness at the half-light radius) correlate in
the same way as elliptical galaxies, yet pseudobulges do not
participate in these correlations. \cite{gadotti2008} shows that many
bulges fall below the Kormendy relation, and are thus lower in surface
density per radial size than a similar sized elliptical galaxy.

As stated above, the dichotomy in bulge properties extends to the ISM
properties of bulges. \cite{regan2001bima} compare the radial
distribution of CO to the stellar light profiles in 15 spiral
galaxies.  They find that most of the galaxies in their sample show an
excess of CO emission in the bulge region of the galaxy, and further
that the central CO radial distribution is similar to that of the
stellar light. \cite{regan2006} shows a similar result with Spitzer
IRAC~8~$\mu$m (PAH) data. \cite{helfer2003} find that 45\% of the
galaxies in the BIMA SONG survey have a peak CO emission within the
central 6\arcsec, while many galaxies have a central hole in the CO
map.  This is similar to the result of \cite{fisher2006}, described
above. Thus their appears to be a bimodal distribution of ISM
properties in nearby bulges.

Stellar populations and age-gradients in bulges of disk galaxies
suggest multiple formation mechanisms as well.  \cite{peletier1996}
show with optical and near-IR colors that many, but not all, bulges
are young. \cite{macarthur2004} find that earlier-type, more luminous,
and higher surface brightness galaxies are older and more metal-rich,
suggesting an early and more rapid star formation history for these
galaxies Recent work with the SAURON survey continues to show such
results. \cite{peletier2007} shows that a large fraction of bulges
fall below the Mg$_2$~-~$\sigma$ correlation of Coma cluster
ellipticals, as do all the bulges in Sb-Sd galaxies in the sample of
\cite{ganda2007}. There is evidence that those bulges in
\cite{falcon2006} with central velocity dispersion drops tend to be
younger. However, \cite{moorthy05} find that bulges in their sample
follow similar correlations of stellar populations and dynamics as
elliptical galaxies. Also, \cite{thomas2006} suggest that only the
late-type bulges in their sample could have been significantly
affected by slow internal growth. As in ISM properties, stellar
populations seem to come in two separate flavors, some bulges are young
and others are old.

Many observations indicate that the properties of pseudobulges are
linked to those of their associated outer disks. Observed connections
between bulge and disk stellar populations
\citep{peletier1996,macarthur2004}, inter-stellar medium
\citep{regan2001bima}, and scale lengths \citep{courteau1996} suggest
that pseudobulges form through processes intimately linked to their
host disks. \cite{fisherdrory2008} show that the connection between
the radial sizes of bulges and disks only exists for
pseudobulges. Similarly, \cite{fisher2006} shows that only
pseudobulges have ISM properties and SFR like their outer
disks. These connections between pseudobulge and disk properties
motivate some authors to consider the possibility that pseudobulge
formation is linked to disk properties.

In summary, the observations suggest that pseudobulges are rotating
rapidly, actively forming stars, and structurally different than
elliptical galaxies. Furthermore, many properties of pseudobulges
(e.g.~radial size, and stellar populations) are linked to their outer
disk. Yet, classical bulges are dominated by random motions, contain
old stars, and are structurally similar to elliptical galaxies;
they're properties thus far appear somewhat independent of the
surrounding disk. 

In this paper we use data from Spitzer Space Telescope, Hubble Space
Telescope, and the Galaxy Evolution Explorer to study the present day
growth of bulges in nearby disk galaxies. We use specific star
formation rates to estimate the time-scales for bulge formation. Also,
we report on connections between star formation rates in bulges and
structural properties of those bulges and with properties of their
associated outer disks.

\section{Implications Of Secular Formation Of Pseudobulges}

To be explicit, in this paper the term ``pseudobulge'' is purely
observational. We classify a bulge as pseudobulge only if the bulge
has disk-like morphology, S\'ersic index lower than two, or both of
these (discussed in more detail in \S 4). Separate from this
observational definition, it has been proposed by many authors that
pseudobulges could have formed through internal disk evolution. Our
aim is to test this hypothesis. If pseudobulges truly form all of
their stellar mass through internal means, the implication would be
that galaxies with pseudobulges are physically more similar to a
bulgeless galaxy.

How can one observe a bulgeless galaxy with $B/T>0$? A large number of
simulations show that non-axisymetries are able to rearrange disk
gas such that the central gas density increases
\citep{simkin1980,combes1981,pfenniger1990,athan92,zhang99}.  If a
hypothetical galaxy initially has a purely exponential stellar mass
density profile ($\Sigma(r)\propto e^{-r}$), but the gaseous inflow
generates a steeper than exponential gas profile, the central star
formation rate density will be enhanced accordingly
\citep{kennicutt98,wu2005}.  If the central few hundred parsecs of
this hypothetical galaxy have a greater SFR per unit mass than the
outer parts do, then eventually the stellar density profile will become
steeper than an exponential profile. When one applies typical
bulge-disk decomposition machinery that assumes an exponential disk
and S\'ersic bulge to observations of the hypothetical galaxy, the
result will be $B/T>0.$ This scenario is typically referred to as
``secular'' bulge growth.

Observationally the hypothetical galaxy has a bulge, but theoretically
speaking its just that this disk galaxy has a stellar density profile
that is steeper than an exponential. Given that we can not know for
certain what happened to make a particular pseudobulge, we choose a
purely observational terminology to label bulges. In what follows we
will not assume {\em a priori} that our classification implies
distinct physical nature.

The evidence suggest that a large fraction of disks are barred and
those bars are long-lived \citep{eskridge2000,jogee2004}. Connections
between the presence of bars and bulge growth give credence to the
secular bulge growth hypothesis. In simulations, barred potentials are
efficient mechanisms to drive gaseous inflows
\citep{athan92}. Observations show that galaxies with bars tend to
have higher molecular gas densities than those without
\citep{sheth2005}.  As well, \cite{gadotti2001} find in a sample of
257 Sbc barred and unbarred systems that blue star-forming bulges are
predominantly in barred galaxies. \cite{sakamoto1999} estimate that
the mean rate of inflow of molecular gas in barred galaxies must be
0.1-1~M$_{\odot}$ yr$^{-1}$.  Furthermore, molecular gas densities and
dynamics in barred galaxies suggest that active star formation is
currently building rapidly rotating bulges \citep{jogee2005}.

\begin{figure}[t!]
   \centering
\includegraphics[width=.5\textwidth]{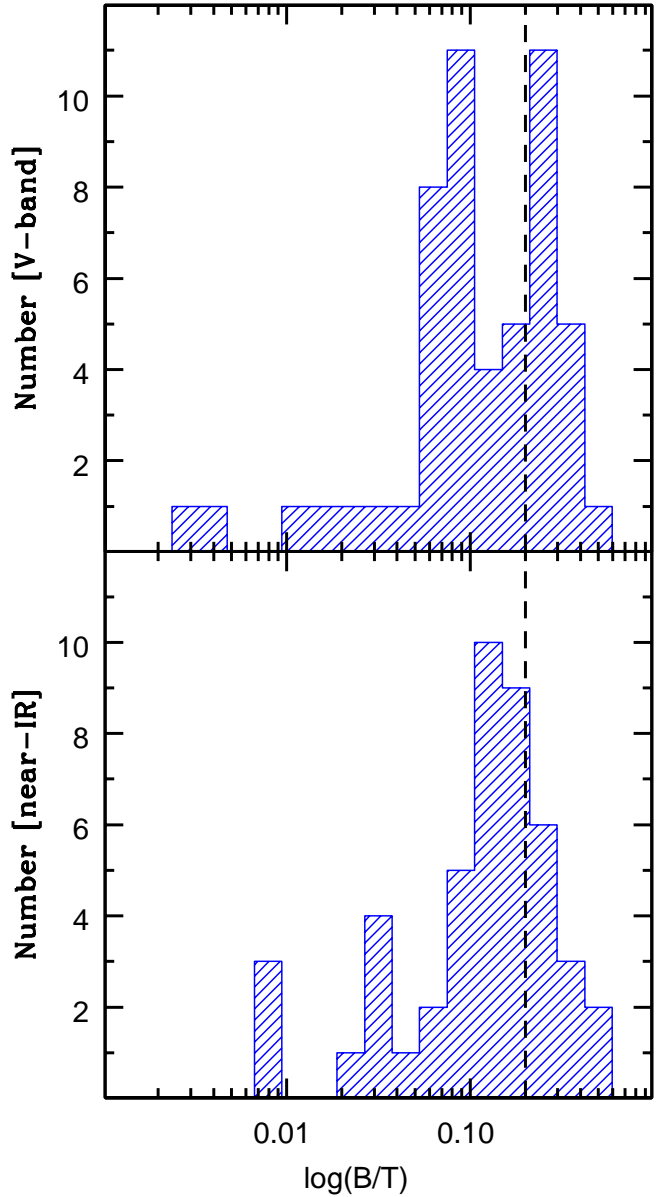}
 \caption{Top Panel: Distribution of $B/T$  of pseudobulges from \cite{fisherdrory2008} in $V$-band. Bottom Panel: The distribution of $B/T$ of pseudobulges from \cite{spitzer1}, in near-IR. 
   \label{fig:b2t}}
 \end{figure}

 Though bulges in barred galaxies on average do have higher SFR, many
 barred galaxies exist with bulges that are not currently growing.
 \cite{fisher2006} shows that the growth is better connected to bulge
 morphology, finding that pseudobulges are growing, but classical
 bulges are not.  \cite{fisherdrory2008} find many unbarred galaxies
 with pseudobulges. Indeed, simulations by \cite{zhang99} indicate
 that spiral structure can drive secular evolution in disk galaxies as
 well. Furthermore, spiral structure is certainly a common phenomenon
 in disk galaxy.  \cite{kormendyfisher2005} argue that secular
 evolution in rotationally supported disks is a natural response to
 local energy minimization, and thus given the opportunity, disks will
 innately drive gas inward.  Bars may be sufficient but not necessary.
 Thus it makes sense that secular bulge growth is common, and it
 appears that even mild non-axisymmetries like spiral structure can
 drive gaseous inflows.

 Recent studies show that galaxy formation models do not predict
 enough disk galaxies with low $B/T$.  \cite{weinzirl2008up} finds
 that the predicted fraction of high mass spirals with a present-day
 $B/T\leqq 0.2$ is a factor of fifteen smaller than the observed
 fraction of high mass spirals with such small bulges. Similarly
 \cite{stewart2008} finds in simulations that almost all giant
 galaxies would have accreted a mass that is larger than the mass of
 the Milky Way disk in the past $\sim$10~Gyr. Given that such an
 encounter is likely to destroy a disk, these results, ``raise serious
 concerns about the survival of thin-disk-dominated galaxies within
 the current paradigm for galaxy formation in a $\Lambda$CDM
 universe.''

 If pseudobulges form secularly, then even pseudobulge galaxies with
 observed B/T$\sim$0.3 are still ``bulgeless'' galaxies.  In
 Fig.~\ref{fig:b2t} we show the distribution of bulge-to-total light
 ratios (B/T) for pseudobulges (blue lines) with $V$-band data (top
 panel) from \cite{fisherdrory2008} and near-IR data (bottom
 panel). The vertical dashed line represents $B/T=0.2$. It is
 certainly true that pseudobulges much more commonly have low
 bulge-to-total ratios. However, 33\% of the pseudobulges in
 \cite{fisherdrory2008} and 24\% of the pseudobulges with near-IR data
 have $B/T>0.2$.  Pseudobulges with $B/T>0.2$ are by no means rare
 among pseudobulges. If pseudobulges form all their mass secularly,
 then the data in Fig.~\ref{fig:b2t} would imply that the number of
 bulgeless galaxies is underestimated, and those gaps between
 observation and theory become more dire.

What else might form pseudobulges? It is often assumed that bulges
formed through successive merging of similar-mass sub-halo objects
early on, and remaining gas that was not involved in the merging
process settles into a gas disk
\citep[e.g.][]{steinmetz95,kauffmann96}. Something similar to this
process may be able to describe the population of classical
bulges. Indeed, \cite{droryfisher2007} find that classical bulges
reside exclusively in red sequence galaxies.  Yet, it would be
difficult for the end products of such processes as roughly equal-mass
merging with violent relaxation to make bulges with cold dynamics and
disk-like morphologies. Furthermore, major-mergers are likely to
consume most of the fuel for future star formation (see
\citealp{schweizer2005} for a recent review). Yet, as discussed above,
cold molecular gas is not-at-all rare in bulges of disk galaxies.

Perhaps successions of minor-mergers with high gas ratios are
responsible for pseudobulges formation. \cite{cox2008} show that the
effect of merging and accretion on the resulting galaxy is a function
of the mass ratio. It is thought, though, that bulge growth by
subsequent accretion of mass results in heating and eventually
destruction of a galactic disk \citep{toth1992,velazquez1999}, and
recent simulations \citep{purcell2008up} also suggest that is hard for
accretion not to destroy a thin disk. 

Other scenarios for the formation of bulges have been suggested. Clump
instabilities in disks at high redshift can form bulge-like structures
in simulations \citep{noguchi1999}. Many recent observations show that
this process may indeed be happening at high redshift
\citep{genzel2008up,bournard2008}.  However, recent work by
\cite{elmegreen2008} suggests that bulges built through clump
instabilities in simulated galaxies better resemble classical bulges.
One should therefore keep in mind that the population of bulges as a
whole, and any one particular bulge, may be the result of more than
one of these processes.

\section{Estimating Time scales For Bulge Growth}

Is the amount of star formation in typical pseudobulges enough to
significantly alter their stellar mass? This is what we seek to
estimate in this paper.  If we assume a continuous gas supply from
internal disk evolution, and approximate that historic SFR as a
constant, we can use the present-day SFRs in bulges to determine the
time-scale for pseudobulge growth.

Mass growth in bulges can be described as
\begin{equation}
M_{star} =M_0+\int_{\tau_{SF}}\psi_{SF}(t)~dt + \int_{\tau_{X}}\psi_X(t)~dt,\label{eq:mass}
\end{equation}
where $M_{star}$ is the current stellar mass; $\psi_{SF}$ is the star
formation rate; $\psi_X$ is the rate at which previously formed stars
are transferred to the bulge (either by accretion or by secular
evolution); $\tau_{SF}$ and $\tau_X$ are the time scales over which
each of these phenomena occur; and finally $M_0$ is that mass that
exists initially in the bulge region. Assuming constant growth (and
that $\tau_{SF}\approx \tau_X\approx t_{grow}$) this can be
simplified, and solved for $t_{grow}$,
\begin{equation}
t_{grow} = \frac{M_{star}-M_0}{\psi_{SF}+\psi_X}.\label{eq:tgrow_init}
\end{equation}
SST provides the ideal instrument to measure $t_{grow}$. Using the
3.6~$\mu m$ luminosity from SST IRAC CH~1 we can measure $M_{star}$,
where $M_{star}/M_{\odot}=L_{3.6}\times(M/L)_{3.6}$ (this is discussed
in more detail below) and the 24~$\mu m$ luminosity obtained with SST
MIPS CH~1 can measure the star formation rate \citep{calzetti2007}.

To measure $M_0$ we subtract the inward extrapolation of an
exponential profile fit to the outer disk. Thus we set
\begin{equation}
M_0= 2\pi \Sigma_0 \int_0^{R_{XS}} r_{\ }e^{r/h}dr \label{eq:mzero},
\end{equation}  
where $\Sigma_0$ is the central mass density of the outer disk, $h$ is
the scale-length of the outer disk, and $R_{XS}$ is the radius at
which the bulge is 25\% brighter than the disk. We set the bulge mass
as
\begin{equation}
M_{XS}\equiv M_{star} - M_0. \label{eq:mxs}
\end{equation}
It is likely that $M_{XS}$, as defined in Eq.~\ref{eq:mxs}, is only a
rough estimate.  Giant disk galaxies may have formed with central
profiles that are cuspier than exponentials. Alternatively, some
galaxies, such as M~104, have central holes in the gaseous disk. Also,
as pseudobulges grow (at times to $B/T\sim1/3$;
\citealp{fisherdrory2008} and Fig.~\ref{fig:b2t}) the structural
parameters of the disk are likely to change. Hence, it may follow that
inward extrapolation of the outer stellar disk of those galaxies may
be inaccurate.  Nonetheless, we feel that $M_{XS}$ is likely a
reasonable estimate of bulge mass for most galaxies.

We make the approximation that star formation internal to the bulge
need only account for a fraction of the mass, hence
\begin{equation}
  t_{grow} \approx \beta \frac{M_{XS}}{\psi_{XS}}, \label{eq:tgrow}
\end{equation}
where $\beta$ is a quantity that measures the amount of stellar mass
growth that is from star formation internal to the bulge, and
$\psi_{XS}$ is the SFR at $z=0$. For the sake of simplicity the
``growth times'' quoted in this paper will assume $\beta=1$ unless
otherwise stated. The $\beta$ factor is likely the product of the
following two phenomena: (1) SFRs that are not constant, and, (2) the
fraction of stellar mass that migrates to the bulge.

The ratio of the the present-day SFR to the average historic SFR
(called the {\em birth-rate parameter}, $b=\psi/\langle \psi(t) \rangle$;
\citealp{scalo1986}) is known two range from $b=0.2-2$ in local disks
galaxies, and is preferentially larger in late-type galaxies
\citep{kennicutt1994}.  If pseudobulges are anything like their outer
disks, then Sa-Sbc pseudobulges are likely to have lower values of $b$
(and thus $\beta$) than pseudobulges in late-type galaxies.  Thus, it
is necessary to consider present day SFRs in intermediate-type
galaxies with different expectations than those in late-type galaxies.

We cannot yet measure the rate at which previously-formed disk stars
are transferred to the bulge. Simulations indicate that it is
occurring, though. \cite{roskar2008} finds that a non-negligible amount
of stellar mass migrates within the disk in simulated disk
galaxies. Further, many pure $N$-body simulations are able to move
mass around within a disk without the presence of any gas
(e.g.~\citealp{pfenniger1990,norman1996,debattista2004}). Also,
\cite{cox2008} show that accretion of relatively low mass galaxies
does not significantly alter a galaxy's SFR.

The total fraction of stellar mass that SFR must therefore account for
is $\beta=b\times(1-\beta_X)$, where $\beta_X$ is the mass that is
transferred as stars formed outside the bulge,
$\beta_X=M_{star}-t_{grow}\psi_X$. There will always be some
degeneracy between $t_{grow}$ and $\beta_X$, in fact it is possible
that the only way we can ever know $\beta_X$ is through simulation,
not observation.

We have no detailed models on which to base our predictions. Yet, we
can place them inside the context of disk-galaxy evolution, and use
what we know about star formation in disk galaxies to make estimates
for how long it might take to form pseudobulges.  Typical SFR
densities in galactic disks are about
$0.01-0.1$~M$_{\sun}$~yr$^{-1}$~kpc$^{-2}$ \citep{kennicutt98}. Bulges
are typically about 1~kpc in radius \cite{spitzer1}. Therefore, if
pseudobulges grow stars at similar rates to the high end of the
distribution for disks, then we expect them to form roughly
$0.1-0.5$~M$_{\sun}$~yr$^{-1}$. If a bulge is $10^9$~M$_{\sun}$ then
it should take a few billion years to make a bulge through internal
disk evolution. Given that disks are not too much older than 5-10~Gyr
\citep{belldejong00}, we expect to find the bulges still forming stars
today. Also, if pseudobulges form too quickly we run into a problem
again, because not every galaxy has a large
pseudobulge. \cite{kautsch2006} find that of giant disk galaxies
roughly 1/10th are ``simple disks'' meaning they have no detectable
bulge when viewed edge-on. We know already that gas consumption time
of the most actively star forming bulges, calculated by \cite{kk04},
tend to be in the fast region of this range, about 1~Gyr.  Thus we
suggest that if the growth time of pseudobulges are significantly
outside the range of 1-10~Gyr, then this would pose a problem for
secular evolution scenario of pseudobulge formation. Furthermore
variations in historic SFR can account for differences of at most a
factor of two outside of this range.

\section{Methods}\label{sec:galaxy-sample}
\subsection{The Sample}

The purpose of this work is to study the present day growth of bulges,
and thus we wish to sample galaxies with a wide range of bulge
properties. Therefore, our sample of 53 galaxies spans the Hubble
types from Sa to Sd.

We begin by visually selecting galaxies from the Carnegie atlas
\citep{carnegieatlas} with distance less than 20~Mpc, such that all
galaxies are at least resolved to a few hundred parsecs with MIPS on
board the SST.  Also, we restrict our sample to exclude significantly
inclined galaxies, we only keep galaxies that satisfy $i$<80\degr.  We
also select galaxies that have ``well behaved'' morphology: free of
tidal-tails, warps and asymmetries to exclude galaxies with significant
interaction-induced star formation.

Though not volume limited, our sample is constructed to cover
parameter space, especially a sequence in mass.  To do this, we select
galaxies covering a range in Hubble types from Sa to Sd; Our sample
consists of 15 Sa - Sab, and 21 Sb - Sbc, 17 Sc-Sd galaxies. Galaxies
in our sample are not fainter than -17 absolute $B$-band magnitude,
and they are typically distributed with $\pm 1$ magnitudes around the
mean of -19.5 $B$-mags.

The link between non-axisymmetries (barred and oval distortions) and
secular evolution motivates us to create a sample containing roughly
equal numbers of galaxies with a driving agent (galaxies with a bar
and/or an oval) and galaxies without a driving agent
\citep{kk04,peeples2006}. Indeed, a correlation between central SFR
and the presence of bars and ovals has been found.
\citep{sheth2005,fisher2006}. Detection of oval distortions are
discussed in \cite{k82}. They are identified by nested shelves in the
surface brightness profile usually having different position
angles. We identify bars by consulting the Carnegie Atlas of Galaxies
\citep{carnegieatlas}, the RC3 \citep{rc3}, and visual identification
in 3.6~$\mu$m images. If a galaxy has both a bar and an oval, we call
that galaxy barred.  Note that we do not distinguish grand design
spirals as a possible secular driver, though they may be able to
generate a similar but less extreme effect as bars do (KK04). In our
sample 22 galaxies are unbarred and unovalled, and 31 are driven (25
barred and 6 ovaled).

\begin{figure*}[t!]
   \centering
 \includegraphics[width=.9\textwidth]{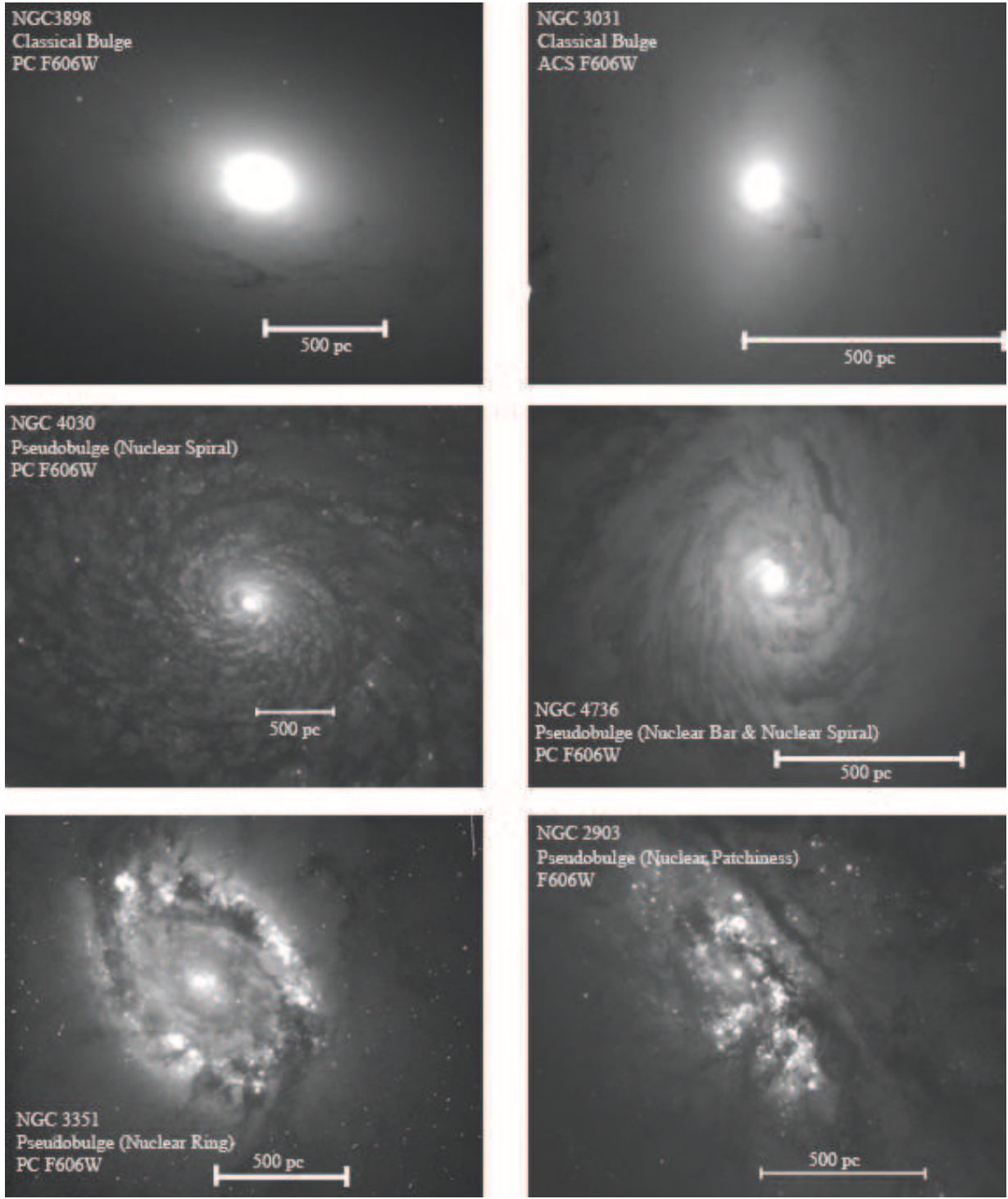}
 \caption{Here we replot a few exemplary galaxies from \cite{fisherdrory2008}. All images are HST F606W images the white line in each image represents 500~pc. The top two rows are examples of classical bulges. The bottom four galaxies are all pseudobulges. 
   \label{fig:bulgeims}}
 \end{figure*}

\subsection{Identification of pseudobulges}\label{sec:ident-pseudo}

In this study, we classify galaxies as having a pseudobulge using two
methods bulge morphology and S\'ersic index.  If the ``bulge'' is or
contains a nuclear bar, nuclear spiral, and/or nuclear ring the
``bulge'' is a pseudobulge. Also, if the bulge has S\'ersic index less
than two, the bulge is called a pseudobulges. Conversely if the bulge
is featureless and has a higher S\'ersic index, the bulge is called a
classical bulge. However, \cite{fisherdrory2008} show that about 10\%
of bulges with S\'ersic index higher than two, have disk-like nuclear
morphology.  For a detailed description of this method, see
\cite{fisherdrory2008}.  Examples of nuclear morphology that indicates
a bulge as a pseudobulge or a classical bulge are shown in
Fig.~\ref{fig:bulgeims}.

Their is significant overlap of our sample with
\cite{fisherdrory2008}, 31 galaxies are in both samples. We therefore
use bulge S\'ersic indices from \cite{fisherdrory2008}, when
available. For the remaining 23 galaxies we generate new bulge-disk
decompositions using archival data from HST archive, Sloan Digital Sky
Survey, and NASA Extragalactic Database (NED). Bulge S\'ersic indices
for all galaxies are given in Table 1. 

Our decomposition method is discussed in the Appendix. We also show
results of each new fit in both figure and tabular form.  The method
we use to calculate surface brightness profiles and S\'ersic fits to
those profiles is the same procedure as used in
\cite{fisherdrory2008}. This procedure is also employed in
\cite{kormendy2008virgo} on elliptical galaxies. We refer interested
readers to these two papers for more detailed discussions of our
reduction and analysis software and procedures.

We identify pseudobulges using HST archival images in the optical
wavelength regime ($B$ through $I$). This makes bulge classification
subject to the effects of dust. However, the structures used to
identify pseudobulges are usually experiencing enhanced star formation
rates, and are easier to detect in the optical region of the spectrum
where the mass-to-light ratios are more affected by young stellar
populations, rather than in the near infrared where the effects of
dust are lesser. Classical bulges may have dust in their center, as do
many elliptical galaxies \citep{Laueretal05}. The presence of dust
alone is not enough to classify a galaxy as containing a
pseudobulge. We indicate which galaxies are pseudobulges, and
classical bulges in Tables 1 \& 2. These structures are often present,
and affect the surface brightness profile, even in the near-IR, at
3.6~$\mu$m, where differences from varying mass-to-light ratios are
minimized. 

We use the NASA Extragalactic Database NED to search for any
evidence of close companions of similar magnitude, tidal distortions,
or peculiar morphology. We remove those galaxies which seem to be
interacting with other galaxies from our sample. Three galaxies in our
sample have companions at $\sim 100$~kpc, which do not appear to
affect the morphology of these galaxies' disks. However, M~51 is a
notable exception to this rule as it is currently accreting a smaller
galaxy.

\subsection{Photometry}\label{sec:photometry}

Imaging data used to calculate fluxes for this paper comes from the
following sources: Spitzer IRAC CH~1 (3.6~$\mu$m), Spitzer MIPS CH~1
(24~$\mu$m), GALEX FUV \citep{galex} and HST NICMOS. The IRAC and
NICMOS images are used to calculate stellar mass (discussed below),
and the GALEX and MIPS images are used to determine SFRs. We use
post-BCD frames for all Spitzer data, and pipeline reduced GALEX and
HST data.

To measure 3.6~$\mu$m surface brightness profiles we use the code of
\cite{bender1987}. First, interfering foreground objects are
identified in each image and masked. Then, isophotes are sampled by
256 points equally spaced in an angle $\theta$ relating to polar angle
by $\tan \theta = a/b\,\tan \phi$, where $\phi$ is the polar angle and
$b / a$ is the axial ratio.  An ellipse is then fitted to each
isophote by least squares. The software determines six parameters for
each ellipse: relative surface brightness, center position, major and
minor axis lengths, and position angle along the major axis. We then
shift the NICMOS F160W images (when available in the archive) to the
same zero point as the IRAC data. The composite profile is NICMOS data
for r<1.22 arcsec, the average of the two profiles when they overlap,
and IRAC 3.6~$\mu$m data at large radii (typically $r > 10$~arcsec).

We note that this procedure assumes a color gradient of zero from
$L$-band to $H$-band in our bulges. This assumption introduces a
source of uncertainty, yet allows for a more complete description of
the stellar mass profile. To quantify this uncertainty we calculate
the entire radial surface brightness profile in $H$-band using NICMOS
and 2MASS data. We then shift that profile to have the same zero point
as the IRAC~3.6~$\mu$m profile, and then calculate the bulge
luminosity, which we call $L_{3.6(H)}$. The difference
$L_{3.6}-L_{3.6(H)}$ is scaled by the fraction of light that comes
from the shifted NICMOS F160W data. This is taken as the error. This
error is typically less than 5\% and rarely larger than errors from
other sources, such as fitting uncertainty. We use NICMOS data because
it is our belief that the high resolution data increases accuracy,
even if precision is compromised slightly.

Prior to measuring the bulge flux of MIPS images we run the images
through a few iterations of the Lucy-Richardson deconvolution routine
in IRAF; we are primarily interested in reducing the effects of the
Airy rings in the MIPS~24~$\mu$m point-spread-function (PSF).  We
construct a PSF from point sources in the image. However, many of our
images do not include enough high signal-to-noise point sources; in
this case we use the theoretical PSF available on the MIPS
web-site. To calculate the surface brightness profile at 24~$\mu$m we
use the PROFILE tool in the the image analysis package VISTA
\citep{lauer1985}. The 24~$\mu$m luminosity is then calculated by
integrating the 2-D surface brightness profile to the bulge radius
($R_{XS}$), determined using the 3.6~$\mu$m profile. Galactic
extinction is considered negligible for 24~$\mu$m images.  Aside from
deconvolution, we carry out the same procedure to measure FUV
luminosities. We calculate the extinction in FUV using the results
from \cite{cardelli1989} and \cite{schlegel}.

For the 24~$\mu$m and FUV profiles, we consider two sources of error
in calculating our luminosities. First, uncertainty in the choice of
$R_{XS}$ leads to errors in the luminosity calculation. We choose
$R_{XS}$ as the radius at which a galaxy is 25\% brighter than the
inward extrapolation of an exponential profile fit to the outer
disk. We construct an error to this by simply integrating the
luminosity to the next resolved points in the profile. Secondly, we
also consider the variance in the image as a source of error. These
two errors are then combined in quadrature to construct the total
error in luminosty. Typically the uncertainty due to $R_{XS}$ heavily
dominates the total error.

\subsection{Calculation Of  Mass From 3.6 $\mu m$ Luminosity}

We assume that the near-infrared light is a good tracer of stellar
mass due to its weak dependence on star formation history
\citep{aaronson1979,rixrieke1993}. In this paper, we calculate stellar
mass by using the relation between mass-to-light ratio ($M/L$) and
color. We assume that $M/L_{3.6} = <(L_k/L_{3.6})>(M/L_K)$ where
$M/L_K$ is determined from optical colors with $B-V$ as in
\cite{belldejong2001}, and take the mean ratio $L_k/L_{3.6}$ from
\cite{dale2007}. We use the $B-V$ color from the RC3 \citep{rc3}; if
the galaxy does not have a $B-V$ in the RC3 we use the value from
\cite{prugniel1998}. We correct these optical colors for Galactic
extinction using data from \cite{schlegel}.  For the calculation of
stellar mass we assume that total colors are good estimates of the
stellar populations of the bulges; this may introduce a source of
uncertainty. However, the color of the bulge has been shown to be very
similar to the color of the outer disk in intermediate type galaxies
\citep{peletier1996}. Thus it is likely a safe assumption.

\subsection{Contamination From Active Galactic Nuclei}

One difficulty in measuring the SFR in bulges of galaxies is that
active galactic nuclei can contribute significantly to the mid-IR flux
in the centers of galaxies. Most of the galaxies in our sample have an
active non-thermal source in their center; what remains is to
determine what the typical effect is and which galaxies are most
heavily affected.

We use IRAC~8~$\mu$m to determine which galaxies have strong nuclear
point sources due to their increased angular resolution. In a few
galaxies in our original sample, over 80\% of the bulge light is
contained within a point source in the 8~$\mu$m images.  We identify
this light as non-thermal by comparing the [OIII]/H$\beta$ and
[NII]/H$\alpha$ line-ratios \citep{ho1997} and exclude these galaxies
from the rest of the study. The excluded galaxies are NGC~1068,
NGC~4258, and NGC~5273.

We find that for the remaining galaxies the point sources typically
make up less than 10\% of the bulge light. This is within the
typical amount of measurement uncertainty so that it is not necessary to
account for contributions from the remaining low-luminosity active
galactic nuclei in the rest of the sample. In \cite{fabriciusinprep}
we directly investigate connections between growth of pseudobulges and
the growth of central active galactic nuclei.

\subsection{Calibration Of Star Formation Rates}

 In optically thick environments, massive young stars heat dust grains
 which re-radiate that light in the IR. Even though newly formed stars
 are easily detected in the UV, even small amounts of internal extinction
 within those galaxies will hamper efforts to measure the SFR only
 using UV light. For this reason, IR emission has and continues to be
 a good indicator of SFRs in most galaxies
 \citep{kennicutt98}. However, different galaxies have differing
 opacities, and this difference can depend on the ages of the stars
 being probed and the amount of star formation
 \citep[e.g.][]{calzetti1994,bell2003,seibert2005}. Also,
 \cite{bossier2007} find UV emission in the absence of IR emission in
 some nearby galaxies, indicating the existence of unobscured young
 stars.  Therefore, we calibrate a new SFR indicator that combines the
 re-emission from warm dust (MIPS 24~$\mu$m) and directly the emission
 from young stars (GALEX FUV 1350-1750 \AA) luminosities,
 \begin{figure}[t!]
   \centering
\includegraphics[width=.5\textwidth]{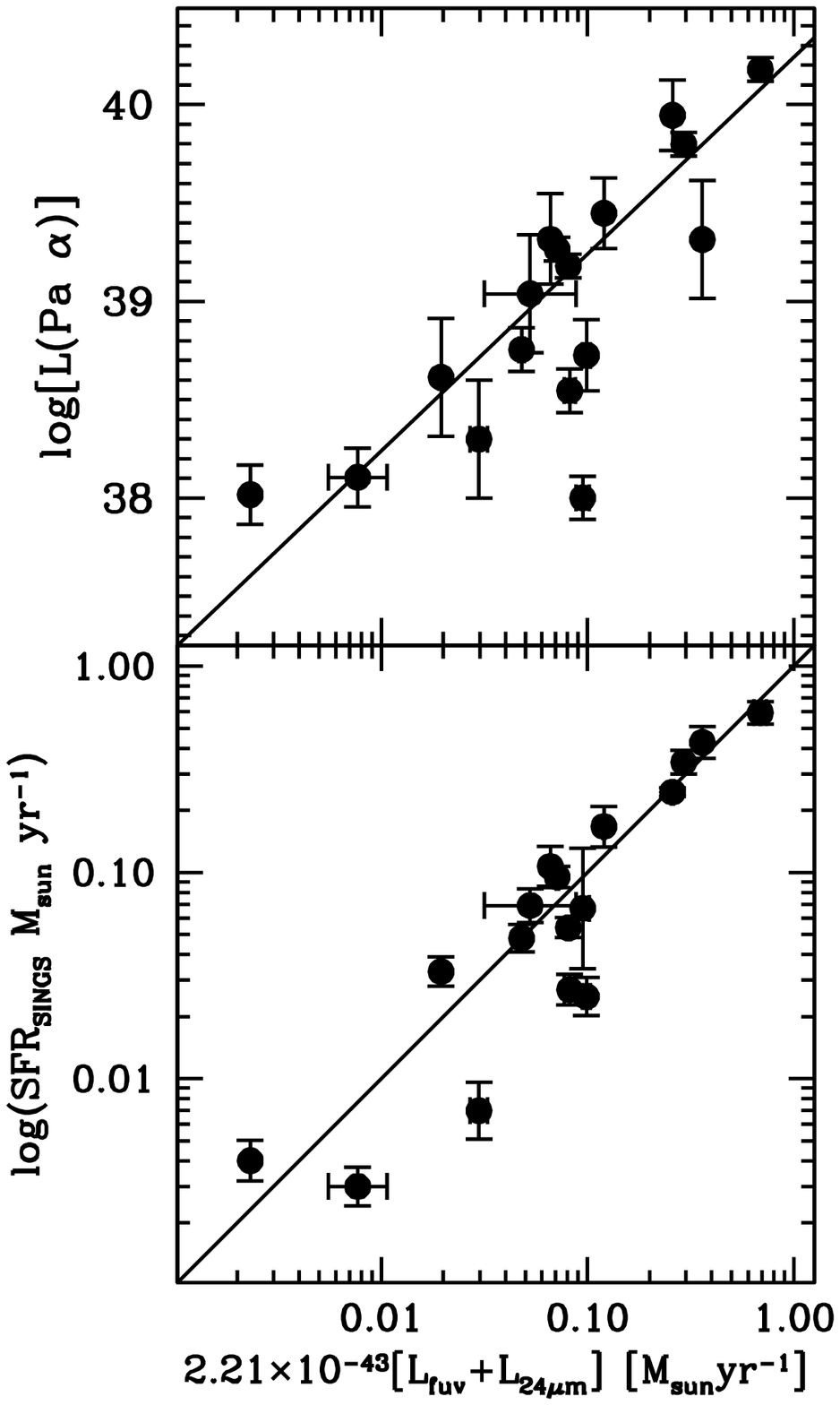}
 \caption{Top Panel: The comparison of our metric of star formation
   rate, $SFR_{UV,IR}=2.21\times0^{-43}[L(FUV) + L(24)]$, to the
   luminosity of Pa $\alpha$ emission. The solid line shows a linear
   relation $L(Pa~\alpha [erg s^{-1}]) =
   1.74\times10^{40}SFR_{UV,IR}[M_{\sun}yr^{-1}]$. Bottom Panel: The
   comparison of our star formation rates to those measured by
   \cite{calzetti2007} the solid line represents the line of
   equality. \label{fig:calib}}
 \end{figure}

 \begin{figure}[t!]
   \centering
 \includegraphics[width=.5\textwidth]{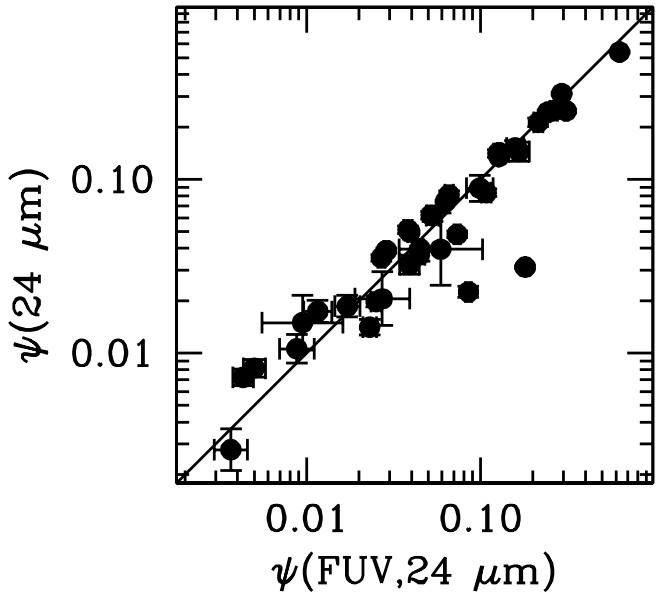}
 \caption{The comparison of the two methods used to calculate SFR in this paper. The solid line represents the line of equality.  
   \label{fig:sfrcomp}}
 \end{figure}

\begin{equation}
  SFR~(M_{\odot}~yr^{-1}) = a\, [L(FUV) + L(24~\mu m)] \label{eq:sfr}
\end{equation}
where $a$ is a conversion constant. A similar SFR indicator is
calibrated by \cite{bigiel2008up}.

We use the ``high metallicity'' galaxies in \cite{calzetti2007} as a
sample on which to calibrate Eq.~\ref{eq:sfr}. These galaxies are
used, in part, because they all have measured Paschen-$\alpha$
luminosity, which is a more direct probe of HII regions, and is much
less affected by internal extinction. Thus, a linear correlation with
the Pa$\alpha$ luminosity would imply that the SFR indicator is
robust. However, there are important distinctions between the methods
used in this paper and those of \cite{calzetti2007}. They measure
luminosity of the central regions of galaxies by summing the
luminosity of individual point sources within those images; we measure
luminosity, as described above, using isosphote measurements. Each
method has advantages and disadvantages, and we do not wish to claim
either is better or worse. Our method will measure a certain amount of
diffuse emission that may not have been counted by
\cite{calzetti2007}, and because we calculate isophotes based on the
mean value of an ellipse, our method may underestimate the luminosity
of extremely bright knots of star formation. We note that these two
effects act against each other, and may lessen systematic
differences. In interest of measuring SFRs that are comparable across
these two methods, we calibrate Eq.~\ref{eq:sfr} using our
measurements of $L(FUV)$ and $L(24~\mu m)$ and those SFRs taken from
\cite{calzetti2007}.  Also, \cite{calzetti2005} measures luminosities
in a $51\times51$ square arcsecond field (the field of view of NICMOS
3); we adjust this to an elliptical aperture that matches our
galaxies assuming a constant azimuthal density. We expect that this
affects the measurements very little, but is a difference
none-the-less.

The constant $a$ is intended to scale our combined luminosities to
units of SFR. We find
\begin{equation}
  a = 2.21\times10^{-43}~M_{\odot}~\mathrm{yr}^{-1}~\mathrm{erg}^{-1}~\mathrm{s}.
\end{equation}
In the top panel of Fig.~\ref{fig:calib}, we show the comparision of
our SFR indicator to the Paschen~$\alpha$ emission. We find good
agreement with a linear correlation (overplotted as a solid line) of
our indicator with Paschen~$\alpha$, thus indicating that our
estimates are robustly measuring the high-mass star formation. In the
bottom panel of Fig.~\ref{fig:calib}, we show the comparison our SFR
to that of \cite{calzetti2007}; a line of equality is overplotted. We
find good agreement between these two different SFR indicators.

Not every galaxy in our sample has GALEX data available. Typical disk
galaxies like the ones in our sample would be classified as ``high
metallicity'' based on the criteria in \cite{calzetti2007}; therefore
single-band fluxes would be considered sufficient. However, if more
data exists that may improve the reliability of our SFRs then we ought
to use that data.  Therefore, when both FUV and Spitzer data are
available we use the SFR indicator described above, and a single band
indicator when only 24~$\mu$m data is available. In our sample 35
galaxies have both 24-$\mu$m and FUV datas, and 17 galaxies only have
24~$\mu$m data.

To measure the SFR from 24~$\mu$m luminosity alone we use all galaxies
in our sample that have both FUV and 24~$\mu$m data.  We find that
single-band 24~$\mu$m-luminosity SFRs (using the calibration from
\citealp{calzetti2007}) are systematically
low compared the the SFR computed with Eq.~\ref{eq:sfr}, although the
exponent of the correlation appears the
same ($SFR\propto (L_{24\mu m})^{0.885}$).  In attempt to account for
this we multiply the equation from \cite{calzetti2007} for 24~$\mu$m
alone by the mean fractional difference, which we find
$<SFR(FUV,24)/SFR(24)> = 1.3\pm0.3$. Thus we use
\begin{equation}
  \psi(24 \mu m) = 1.65\times10^{-38}(L_{24~\mu~m})^{0.885}, \label{eq:irsf}
\end{equation}
where $L_{24~\mu~m}$ is in ergs~s$^{-1}$, as opposed to the original
formula which has a multiplier of 1.27. This same scaling difference
exists when comparing our single flux measurements to the SFR in
\cite{calzetti2007} for those galaxies that are present in both
samples. This small difference in scaling is likely a concequence of
the different methods to calculate the bulge luminosity.  Our method
intergrates azimuthaliy averaged isophotes, which likely reduces the
effects of bright sources.

In Fig.~\ref{fig:sfrcomp} we compare SFR calculated with both methods,
our single-band SFR and our FUV plus 24~$\mu$m indicator. As one can
see from Fig.~\ref{fig:sfrcomp} 33 of 35 bulges have very similar SFR
as measured by IR+FUV indicator or the single-band IR indicator
(standard deviation of the difference between the two indicators is
0.02~M${\odot}$~yr$^{-1}$). The two outlying galaxies (NGC~0925 \&
NGC~1512) show an unusually large number HII regions
\citep{carnegieatlas}. We check NED for similar comments on all our
galaxies that contain only IR data, of those 17 only NGC~3726 has
similar comments.  We conclude in general single band fluxes yield a
good estimate of the SFR, and in rare circumstances, likely requiring
unusually high numbers of HII regions, single band IR calibrations may
understimate the SFR.  Those galaxies that do not have GALEX
observations are indicated in Table~2. Also, in
Fig.~\ref{fig:sfr_methods} we replot our principle result such that
symbols indicate the different methods used to determine SFR.

 \begin{figure}[t!]
   \centering
 \includegraphics[width=.5\textwidth]{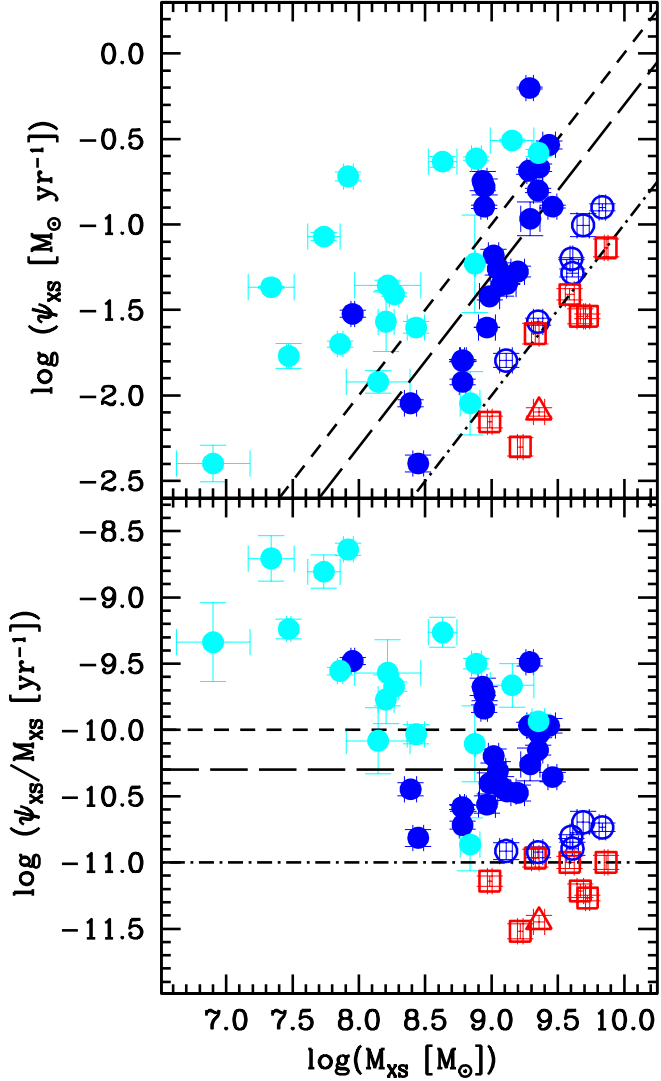}
 \caption{Top Panel:Dependence of star formation rate on stellar mass
   of the bulge (as defined as the excess mass above the inward
   extrapolation of the exponential disk). Bottom Panel: Specific star
   formation rate (star formation rate per unit mass) plotted against
   bulge mass. The lines indicate from top to bottom
   $t_{grow}=10,20,100$ Gyr, short dashes, long dashes, and dot-dashes
   respectively. In both panels, and all figures here after, the
   symbols are as follows: pseudobulges are indicated by filled blue
   circles, centers of late-type disks by green x's, inactive
   pseudobulges are denoted by blue open circles, and classical bulges
   are denoted by red open squares. In each figure we denote M~81 as
   an open red triangle. \label{fig:sfr_xs}}
 \end{figure}

 \begin{figure}[t!]
   \centering
 \includegraphics[width=.5\textwidth]{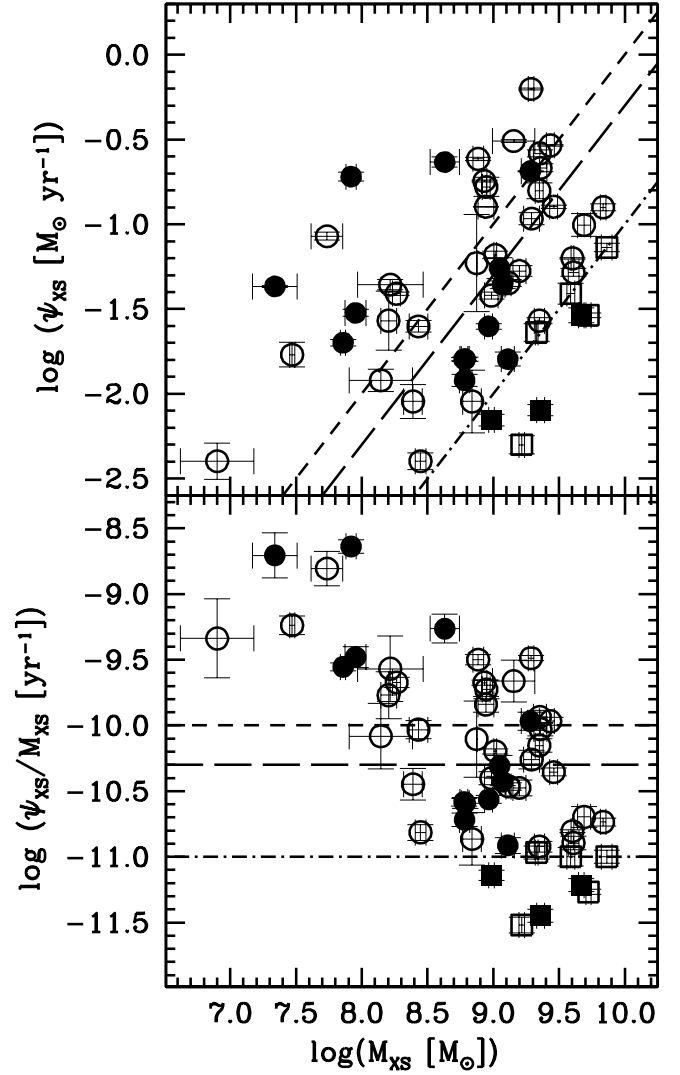}
 \caption{This figure is the same as Fig.~\ref{fig:sfr_xs}, except here we distinguish galaxies based on the method used to calculate the SFR. Bulges using 24 $\mu$m are represented by filled symbols, open symbols represent those bulges that use both FUV and 24 $\mu$m to determine the SFR. As in Fig.~\ref{fig:sfr_xs} pseudobulges are represented by circles, and classical bulges are represented by squares. 
   \label{fig:sfr_methods}}
 \end{figure}

\section{The Growth Of Pseudobulges In Galactic Disks}\label{sec:sfr}

\subsection{Growth Times In Pseudobulges}

All bulges in our sample are forming some stars, irrespective of
whether they are classical bulges or pseudobulges. This is apparent in
the top panel of Fig.~\ref{fig:sfr_xs}, where we plot SFR of the
bulge, $\psi_{XS}$, versus bulge mass, $M_{XS}$. Typical star
formation rates in bulges range from 0.01 to 1.0 $M_{\odot}$
yr$^{-1}$; both the highest and lowest SFR bulges are
pseudobulges. Generically speaking, SFR of bulges are consistent with
a linear correlation with mass ($\psi_{XS}\propto$M$_{XS}$), where
$\psi_{XS}$ is the total SFR within the bulge radius, $R_{XS}$. The
lines indicate three linear growth models, $\psi_{XS}=M_{xs}/t_{grow}$
where $t_{grow}=10,20,100$~Gyr, represented by dotted lines, short
dashes, and long dashed lines, respectively.  The classical bulge with
the highest SFR is M~81 with $\psi_{XS}=0.65$, denoted as a triangle
in Fig.~\ref{fig:sfr_xs}. M~81 is known to be interacting with nearby
M~82.

We are principally interested in determining if an extended SFR
roughly equivalent to the present-day SFR is able to account for the
growth of the stellar mass in pseudobulges. To better illustrate this
result, in the bottom of Fig.~\ref{fig:sfr_xs} we plot the specific
SFR against bulge mass. The black lines indicate the time necessary to
grow the stellar mass in $t=2,10,20$~Gyr from top to bottom.

We distinguish four types of bulges in this paper. The first
distinction comes from morphology of the bulge and S\'ersic index in
optical bands, as discussed above: classical bulges (open red squares)
and pseudobulges (light and dark blue circles). Further, we
distinguish three types of pseudobulges: pseudobulges in late-type
galaxies (light blue filled circles); active pseudobulges in
intermediate-type galaxies (blue filled circles); inactive
pseudobulges in intermediate-tyep galaxies (specifc SFR
$SFR/M_{XS}<20$~Gyr$^{-1}$ and $M_{XS}\geqq 300 \times 10^6M_{\odot}$,
open blue circles). These four sets of galaxies produce four roughly
parallel sequences in the $\psi_{XS}-M_{XS}$ plane, each growing
roughly linear, and each being offset toward higher mass per unit SFR
as one goes from late-type pseudobulges to intermediate-type
pseudobulges to inactive pseudobulges to classical bulges.

From the figure it is clear that present day SFR in almost all active
pseudobulges is sufficient to account for the mass of those bulges (in
both late and intermediate type galaxies), but not enough to account
for the mass of any of the classical bulges or inactive pseudobulges.  There
is a high-scatter negative correlation between specific SFR of all
bulges with stellar mass that is roughly consistent with mass growth
via constant SFR.
\begin{figure*}[t!]
  \centering
\includegraphics[width=.9\textwidth]{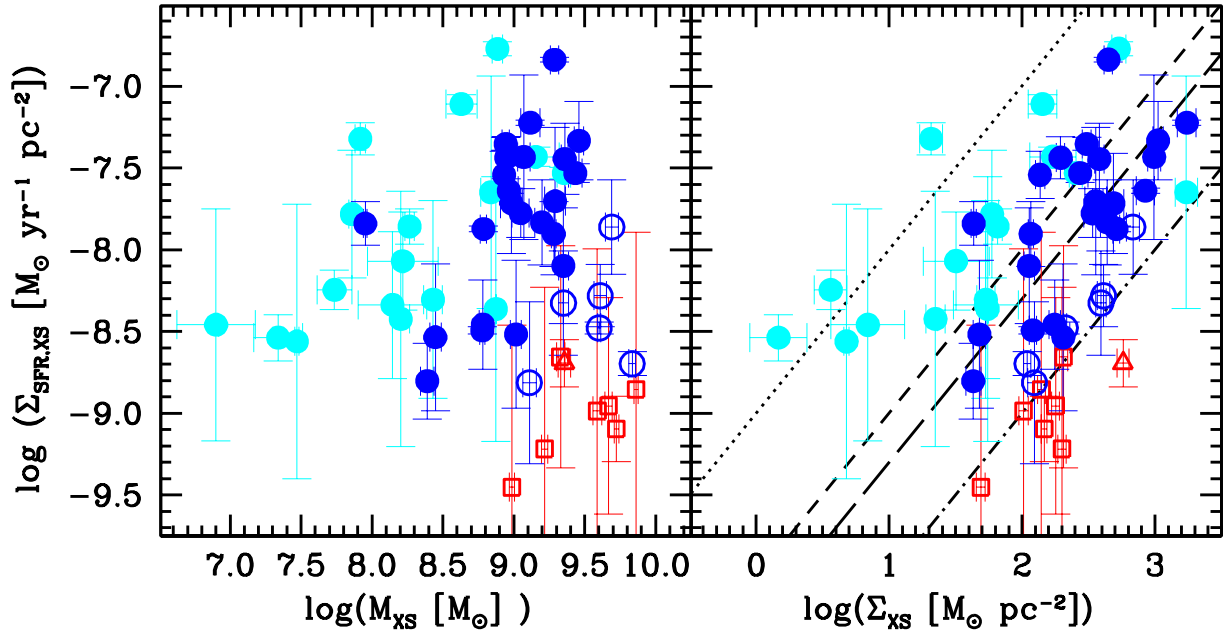}
\caption{Here we show the star formation rate density of bulges as a
  function of bulge mass (left panel) and bulge mass density (right
  panel). The dashed line represents a linear bisecting correlation
  between SFR density and mass density, $\Sigma_{SFR,XS}\propto
  \Sigma_{XS}$, for the (active) pseudobulges only, the solid lines
  represent plus-and-minus one standard deviation around the bisecting
  line. Symbols are the same as
  Fig.~\ref{fig:sfr_xs}.\label{fig:sigsfr}}
\end{figure*}

For all pseudobulges (all 3 types) the median growth time is 12.4~Gyr,
however there is signifigant spread. Of all 45 pseudobulges in our
sample 33 have growth times less than 10~Gyr. For active pseudobulges
in intermediate-type galaxies we find the median growth time is
6.3~Gyr, and 19 of the 39 active-pseudobulges have growth times less
than 10~Gyr, and 30 of 39 would require less than 20~Gyr. The star
formation growth times calculated by \cite{kk04} are on the high end
of this distribution, however we note that they restrict their sample
to nuclear rings, which are forming stars much more vigoursly than the
typical pseudobulge.

We remind the reader that a degeneracy exists between the time
necessary to grow a structure and the fraction of mass for which the
present day SFR needs to account. It is very likely that this ratio
($\beta$, see Eq.~\ref{eq:tgrow} and subsequent discussion) varies
from galaxy to galaxy, and may possibly correlate with mass. Therefore
discussion of a single metric of growth time for all pseudobulges is
likely an oversimplification. Furthermore, we remind the reader that
in typical disk galaxies historic SFR were higher than present day SFR
by roughly a factor of two \citep{kennicutt1994}.

Late-type bulges have the highest specific SFRs due to their small
masses. The mean specific SFR of late-type bulges is
$(\psi_{XS}/M_{XS})_{late}=2.7\times10^{-10}$ yr$^{-1}$. The mean SFR
of late-type bulges is 0.1~M$_{\odot}$ yr$^{-1}$, and the average mass
is $4.4\times10^8~M_{\odot}$. Thus if the mean late-type bulge is able
to maintain a constant SFR for the next gigayear, the resulting bulge
would fall near the low-mass end of the present-day intermediate-type
pseudobulge sequence in Fig.~\ref{fig:sfr_xs}.

Classical bulges and inactive pseudobulges are uniformly not forming
stars at high enough rates to form their stellar masses within a
reasonable amount of time, including M~81. The mean growth time for
classical bulges in our sample is $1.7\times10^{11}$~yr.  Inactive
pseudobulges are slightly higher in specific SFR than classical
bulges.

In Fig.~\ref{fig:sfr_methods} we replot our main result, shown in
Fig.~\ref{fig:sfr_xs}. However here plotted symbols reflect the type
of method used to calculate the SFR; open symbols represent SFR
determined with FUV \& 24~$\mu$m, and solid symbols represent SFR
determined with 24~$\mu$m only. There does not appear to be any strong
bias between the two methods. We reiterate our earlier statement that
single band (24 $\mu$m) fluxes are sufficient to determine the SFR;
however, additional information from UV data improves reliability and
should therefore be included when possible.

\begin{figure*}[t]
  \centering
\includegraphics[width=.9\textwidth]{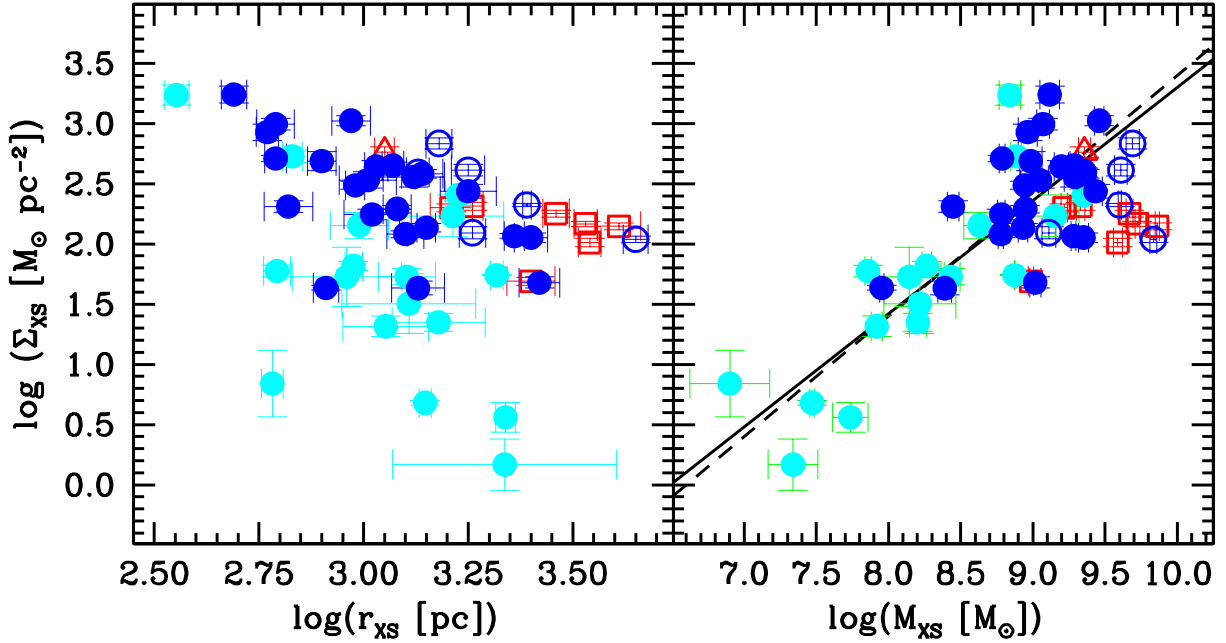}
\caption{Here we show the structural parameters of different bulge
  types. The left panel shows mass density ($\Sigma_{XS}$) versus
  bulge size ($r_{XS}$). The right panel shows mass density plotted
  against bulge mass ($M_{XS}$). Symbols are the same as
  Fig.~\ref{fig:sfr_xs}.\label{fig:struct}}
\end{figure*}

\subsection{A Link Between the Growth of Pseudobulges and their Structure}

In Fig,~\ref{fig:sigsfr} we show the SFR density of the bulge,
$\Sigma_{SFR,XS}$, as a function of bulge mass, $M_{XS}$, (left panel)
and mass surface density of the bulge, $\Sigma_{XS}$ (right panel). To
calculate densities we set $\Sigma_{SFR,XS}= \psi_{XS} /(\pi R_{XS}^2)$, and
likewise for mass density.  The most striking feature in both of these
panels is the absence of low-mass bulges with high SFR densities.

We find that, similar to specific SFR in Fig.~\ref{fig:sfr_xs},
classical bulges have small $\Sigma_{SFR,XS}$ compared to their
masses, and they do not show much correlation between
$\Sigma_{SFR,XS}$ and mass. As was the case with mass in
Fig.~\ref{fig:sfr_xs}, normalizing classical bulges by area shows that
their SFR is insignifigant, especially in comparison to the
pseudobulges. Indeed all bulges are forming stars at similar rates,
yet classical bulges are just too large for the present day SFR to be
important.  Inactive pseudobulges are between the classical bulges and
pseudobulges. We cannot say whether there is any correlation in this
space for the inactive pseudobulges. Finally, late-type bulges have
comparable SFR densities as pseudobulges but are offset to lower mass.

In the right panel of Fig.~\ref{fig:sigsfr}, we show the SFR density
versus mass density of the bulge, $\Sigma_{SFR,XS}$~vs.~$\Sigma_{XS}$.
A weak positive correlation exists between mass density and SFR density of
all pseudobulges. In the figure (Fig.~\ref{fig:sigsfr}) we overplot four lines that indicate different characteristic times in the simple formula, 
\begin{equation}
\Sigma_{SFR,XS}=\frac{\Sigma_{XS}}{t_{grow}},\label{eq:sigsfr}
\end{equation} 
where $t_{grow}=1,\ 10,\ 20,\&\ 100$~Gyr (dotted, short dash, long dash, \&
dot-dashed).  

A linear regression fit to all active pseudobulges shows a weak
postive correlation that is
$\Sigma_{SFR,XS}=10^{-8.79\pm0.08}\Sigma_{XS}^{0.46\pm0.04}$; yet the
correlation has somewhat high scatter, with Pearson correlation
coefficient $r=0.6$. A regression fit to all pseudobulges yields a
similar correlation, but with higher scatter.  In this plane, it
appears as if the inactive pseudobulges lie at the low SFR end of the
correlation defined by the active pseudobulges.  Notice that the power
of the correlation is significantly below unity. This may simply
reflect different times over which significant bulge growth has
occured.  Also, if the ratio of present day SFR to historic SFR
differs systematically from pseudobulge-to-pseudobulge then this
could cause the slope in the fitted correlation to be less steep than
unity.

In Fig.~\ref{fig:struct} we show structural parameter correlations
between bulge surface density $\Sigma_{XS}$ and radial bulge size
($r_{XS}$; left panel), and bulge mass ($M_{XS}$; right
panel). Similar to the right panel of Fig.~\ref{fig:sigsfr} we show
two correlations. Similar results are obtained using the half-light
radius of bulges instead of $r_{XS}$ \citep{spitzer1}. However, the
half-light radius is ill-determined in the absence of high-resolution
data from NICMOS, and using it would therefore restrict us to a much
smaller sample. Further, it seems logical that the radius at which a
disk ceases to be exponential (from outside-in) would be a relevant
metric if secular evolution is occuring within the galaxy.

Pseudobulges show a postive correlation in the mass-density plane. For
all active pseudobulges we find
$\Sigma_{XS}=10^{-6.1\pm0.4}M_{XS}^{0.94\pm0.04}$, with correlation
coefficient $r=0.8$; this is shown as the solid line in
Fig.~\ref{fig:struct}. This correlation is remarkably close to
unity. For comparison we also show a linear bisector
$\Sigma_{XS}=10^{-6.7\pm0.4}M_{XS}$; the bisector is represented by a
dashed line in Fig.~\ref{fig:struct}. The left panel of
Fig.~\ref{fig:struct} shows that when considering all actively growing
bulges, there is no real correlation between surface density of the
bulge and radial bulge size. Thus the radial extent of pseudobulges is
independent of the mass of the bulge. These two results fit well
together. If the radial size of pseudobulges is not affected by an
increase in mass, then surface density and mass ought to have a linear
correlation; indeed, this is what we find.

Inactive pseudobulges have systematically higher density per unit size
than pseudobulges and late-type bulges, and systematically lower
density for a given mass than pseudobulges. In fact, they are in the
same location as classical bulges in both of these plots.

The simplistic assumption that bulges maintain a roughly constant SFR
results in horizontal evolution of pseudobulges in the mass-SFR
density plane (Fig.~\ref{fig:sigsfr}, left panel). Those bulges with
higher $\Sigma_{SFR,XS}$ would move faster, thus vacating the
high-$\Sigma_{SFR,XS}$ low mass region of Fig.~\ref{fig:sigsfr}. As
this growth occurs, the bulge maintains roughly the same radial size,
and thus moves vertically, upward, in the left panel of
Fig.~\ref{fig:struct}. Therefore, as bulges move horizontally from
left-to-right in the $\Sigma_{SFR,XS}-M_{XS}$ plane, they move
diagonally in the mass-density plane with $\Sigma_{XS}\propto M_{XS}$.



\subsection{Connections between The Growth Of Bulges To Outer Disks}

All bulges are forming stars, as was shown in fig.~\ref{fig:sfr_xs},
and SFRs in pseudobulges are high enough to suggest that this mode of
growth contributes a significant fraction of their stellar mass. Now
we wish to know, firstly, if this growth leads to an increase in
stellar mass $B/T$, and, secondly, if this growth is connected to
properties of the outer disk.

\begin{figure}[t!]
  \centering
\includegraphics[width=.5\textwidth]{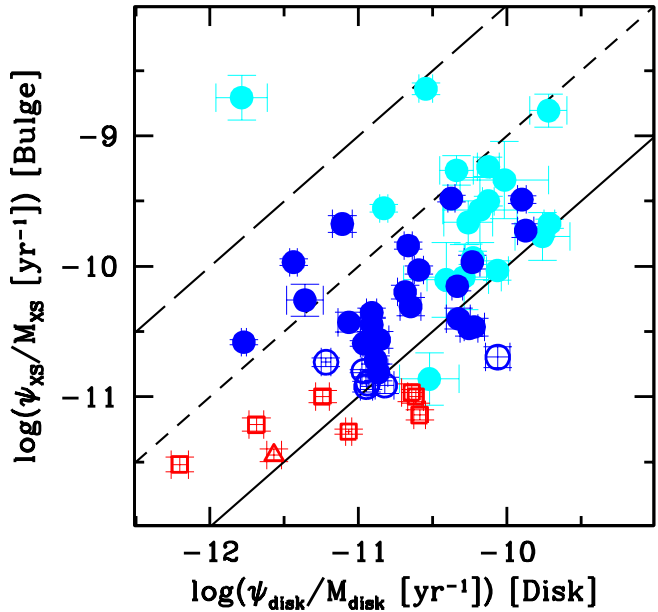}
\caption{Here we show specific SFR of the bulge to that of associated
  outer disk. The solid black line represents the line of equality,
  the line with short dashes represents bulge growth that is 10 times
  that of the disk, and the line with long dashesrepresents bulge
  growth that is 100 times that of the disk. Symbols are the same as
  Fig.~\ref{fig:sfr_xs}. \label{fig:ssfr_diskbulge}}
\end{figure}

If the outer disk were forming stars at high enough rates, then the
galaxy will not increase stellar $B/T$ despite the bulge SFR. The
change in the ratio of bulge mass to disk mass, $M_{bulge}/M_{disk}$,
can be expressed as
\begin{equation}
  \label{eq:b2d}
  \frac{\mathrm{d}}{\mathrm{d}t}\left( \frac{M_{bulge}}{M_{disk}} \right) \, = \, 
  \frac{M_{bulge}}{M_{disk}} \left( \frac{\psi_{bulge}}{M_{bulge}} - \frac{\psi_{disk}}{M_{disk}} \right).
\end{equation}
As one can see from Eq.~\ref{eq:b2d}, the trend in $B/T$ can be
determined by comparing the specific SFR of bulges to that of
disks. If the bulge has a higher specific SFR than the disk, then over
a time the galaxy will evolve toward earlier Hubble types. 

We compare the specific SFR of the bulge to that of the outer disk in
Fig.~\ref{fig:ssfr_diskbulge}.  The solid line in represents the line
of equality, the shortdashed line represents bulge growth that is ten
times that of the disk, and the long-dashed line represents bulge
growth that is one hundred times that of the outer disk. Galaxies
above the solid line in Fig.~\ref{fig:ssfr_diskbulge} are increasing
$B/T$.  We find that almost all ($\sim$80\%) of the galaxies in our
sample are increasing the $B/T$ ratio, and thus evolving toward
earlier Hubble types. Aside from one galaxy that has grown its bulge
extremely fast compared to the outer disk (NGC~4580), the typical
bulge is is growing at 2-6 times that of the outer disk. The small
bulges, pseudobulges in late-type galaxies, are growing much faster
than their outer disk, on average in late-type galaxies the bulge is
growing at a rate roughly eight times that of the outer disk $
<\psi_{XS}/M_{xs}>_{Sc-d} = 8\times<\psi_{disk}/M_{disk}>_{Sc-d}$
(again excluding NGC~4580 from that average).
\begin{figure}[t!]
  \centering
\includegraphics[width=.5\textwidth]{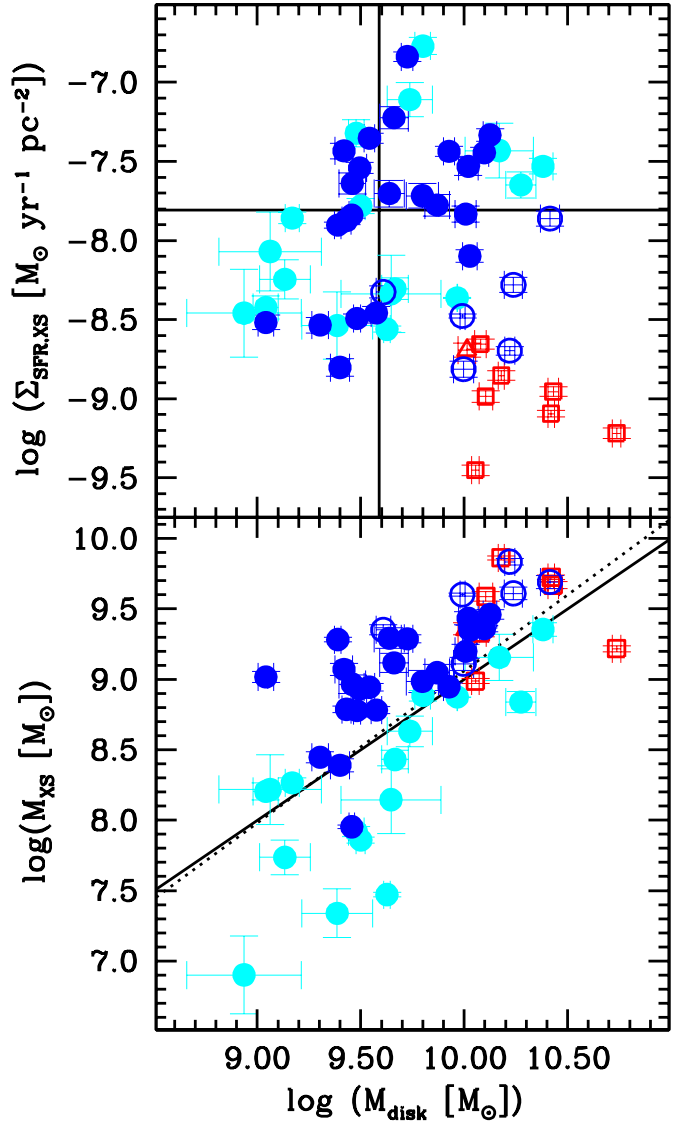}
\caption{Here we compare the specific SFR of the bulge (top) and the
  mass of the bulge (bottom) to the mass of the outer disk.  Symbols
  are the same as Fig.~\ref{fig:sfr_xs}. \label{fig:diskmass}}
\end{figure}

If a galaxy has a classical bulge, the entire galaxy is forming fewer
stars.  \cite{droryfisher2007} show that galaxies with classical
bulges are on the red sequence (as defined by \cite{strateva2001}) and
galaxies with pseudobulges are in the blue cloud. Also
\cite{peletier1996} find that bulge and disk ages are correlated from
galaxy to galaxy, older bulges are in older disks. Similar to these
results, we find in Fig.~\ref{fig:ssfr_diskbulge} that those disks
that are not forming many stars have bulges that are not growing
either, and the bulges that are the most active are in the most active
disks. Thus the present-day SFR in classical bulges will not produce
significant evolution in those galaxies. However, if galaxies with
pseudobulges are able to supply enough gas to support their star
formation, they will evolve considerably.

In Fig.~\ref{fig:diskmass}, we compare the mass of disks to the mass
of the bulges (bottom panel), and to the the SFR density of the bulges
(top panel). Pseudobulges show a high-scatter positive correlation
between the SFR density and the disk mass. In the top panel of
Fig.~\ref{fig:diskmass}, we show a horizontal line representing the
median bulge SFR density (median $\Sigma_{XS,SFR} = 8.3\times
10^{-8}$~M$_{\odot}$~yr$^{-1}$~kpc$^{-2}$) and the vertical line
represents the median disk mass (median
$M_{disk}=2.39\times10^9$~M$_{\odot}$).  Larger disks are driving
higher SFR densities in their centers, if the center of the galaxy
contains an active pseudobulge. Less massive disks do not contain
bulges with high SFR densities.  Pseudobules in late-type galaxies are
located in the same region of the $\Sigma_{SFR,XS}-M_{disk}$ parameter
space as those in intermedate-type galaxies.  In both intermediate-
and late-type objects, if the disk mass is small the bulge is not
forming stars as vigoroursly.

In general we find that larger bulges are in larger disks. This is not
necessarily due to internal evolution, but rather it could be that all
substructure is larger if more mass is available in the halo. Though,
the low number of classical bulges and inactive pseudobulges in our
sample prevent us from saying too much about them. The classical
bulges and inactive pseudobulges in this paper tend to be slightly
more massive per unit disk mass.

Pseudobulges in late-type galaxies are systematically lower in bulge
mass per unit disk mass compared to pseudobulges. Of course, $B/T$ is
part of the definition of Hubble type, so this is in no way suprising.
The boundary seperating late-type and intermediate type galaxies in
our sample is around $M_{XS}/M_{disk} \sim 0.1$, indicated bt the
solid line in the bottom panel of Fig.~\ref{fig:diskmass}. There is
also a fitted relation to all active pseudobulges that indicates that
the correlation of disk mass and bulge mass is almost exactly linear
$M_{XS}\propto M_{disk}^{1.08\pm0.07}$, with correlation coefficient
$r=0.6$. This fit is plotted as a dotted line in
Fig.~\ref{fig:diskmass}.  The spread in the correlation between bulge
and disk mass becomes much greater at low-disk mass.

\section{Discussion}

\subsection{Summary of Results}

In this paper we study the star formation and stellar masses in the
centers of bulge-disk galaxies, with a specific emphasis on
pseudobulges. Primarily, we wish to know if the present day star
formation rate in pseudobulges is sufficient to have played a major
role in the formation of bulges we see today. Amoung those
pseudobulges with presently active star formation, the answer to this
question appears to be yes.  In large pseudobulges (Sa-Sbc) the
present day SFR can account for half the stellar mass in 6~Gyr; in
smaller pseudobulges (Sc-Sd) present day SFR needs only 2~Gyr to form
their entire stellar mass.

In pseudobulges, SFR density postively correlates with both mass and
mass density. A regression fit to all presently active pseudobulges
yields $\Sigma_{SFR,XS}=10^{-8.70\pm0.07}\Sigma_{XS}^{0.46\pm0.04}$.
We argue that if the present-day SFR has been sustained for some time,
then the postive correlations between mass and mass density with SFR
density are expected: bulges with higher SFR densities grow faster;
over a long time this process will evacuate the low-mass
high-SFR-density region of parameter space, as we observe in
Fig.~\ref{fig:sigsfr}. Therefore, positive correlations with mass and
SFR density are constsitent with long-term internal bulge growth.  We
note that this arguement is only valid if pseudobulges do not change
radius as they increase stellar mass; this is indeed consitent with the
observation of \cite{spitzer1} that low-mass pseudobulges are the same
size as high mass pseudobulges.

We investigate the location of inactive pseudobulges in structural
parameter correlations. We often find that inactive pseudobulges are
more similar in these parameter spaces to classical bulges, than they
are to pseudobulges that are actively forming stars.

We find that bulges with higher specific SFR live inside disks with
higher specific SFR, though most bulges are in increasing their
relative mass faster than their outer disk. Therefore, the $B/T$ of
almost all of the galaxies in our sample is increasing. More massive
disks are shown to contain both higher star formation rate densities,
and more massive bulges.

\subsection{Is Secular Evolution Evolution Building Pseudobulges?}

Is secular evolution responsible for building pseudobulges in disk
galaxies we observe today? Many observations of disk galaxies,
combined with results of simulations, strongly suggest that the
rearrangement of disk mass into rings and bars is also funneling gas
and stars to the center of the galaxy (see \citealp{kk04,athan05} and
references therein for reviews).  Although detailed predictions about
the growth of pseuodbulges in disk galaxies do not exist, our results
are consistent with expectations that derive from the idea that
pseudobulges are built out of disk material.

We find a picture emerging from our data that is consistent with
secular growth of bulges in disk galaxies. The specific SFR of bulges
in our sample indicate that the typical pseudobulge requires roughly
5~Gyr to form at the present day SFR (shown in
Fig.~\ref{fig:sfr_xs}). If long-term, moderate SFR was responsible for
evolving galaxies from little-to-no $B/T$ to $B/T\sim1/3$, then we
should not find pseudobulges with low mass and high SFR
density. Indeed this is what we find in
Fig.~\ref{fig:sigsfr}. \cite{spitzer1} show that low-mass pseuodbulges
cover the same radial extent as high-mass pseudobulges; this is
replotted in Fig.~\ref{fig:struct}. If pseudobulges are made slowly
through internal star formation there is no violent event that
rearranges the orbits of stars. Thus these bulges would stay the same
size as they increase their mass, and unlike in elliptical galaxies
and classical bulges the mass density would positively increase with
mass. This is what is found by \cite{spitzer1}.  Finally, if bulges
are forming out of disk material it is reasonable to expect that
larger disks would make larger bulges. This is indeed what we find,
and show in Fig.~\ref{fig:diskmass}. More massive pseudobulges are in
more massive disks, also the highest SFR densities only occur in more
massive disks. Our data suggest that small disks cannot grow large
bulges.

The correlations of bulge SFR density and stellar mass with disk mass
fit in well with other correlations of bulge and disk properties.
There is a well-known correlation between the size of the bulge (as
measured by scale-length or half-light radius) and the scale-length of
the outer disk \citep{courteau96,macarthur2003}.
\cite{fisherdrory2008} show that this correlation only exists in
pseudobulges.  Additionally, \cite{spitzer1} show that at 3.6~$\mu$m
the size of the pseudobulges ($r_{XS}$) and the half-light radius of
the associated outer disk are similarly correlated.  Also,
\cite{carollo2007} show that the mass of bulges is correlated with the
total mass of galaxies. However, this is not too suprising since
bulges contribute significant fractions of the mass.  In
fig.~\ref{fig:diskmass}, we show that bulges are correlated with the
disk mass.  Though this does not really rule out other possible
mechanisms of pseudobulge formation, it seems reasonable that secular
growth of bulges would produce such connections between the stellar
mass of bulges and disks.

\begin{figure}[t!]
  \centering
\includegraphics[width=.5\textwidth]{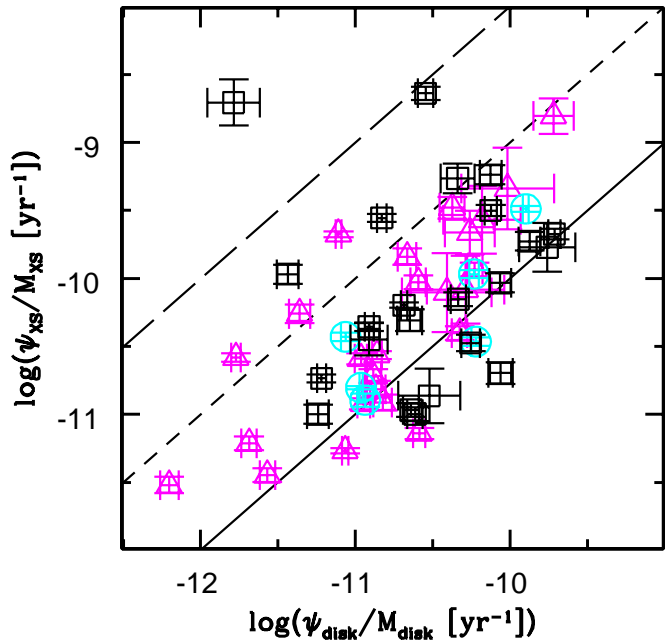}
\caption{Here we replot Fig.~\ref{fig:ssfr_diskbulge}, but the symbols
  are changed to reflect the type of disk each galaxy has. Bars are
  represented by magenta triangles, ovaled disk are represented by
  cyan circles, and galaxies with disks that are neither barred nor
  ovaled are represented by black squares. The solid line shows the
  line of equality, the dashed line shows the case where the bulge is
  growing ten times faster than the disk.\label{fig:drivers}}
\end{figure}

We find the correlation between the specific star formation rates of
the disk and bulge particularly compelling for secular evolution. That
the specific SFR of bulges and disks are correlated is no surprise;
correlations between the stellar populations of bulges and disks are
well known \citep[e.g.][]{peletier1996}. What we show in
Fig.~\ref{fig:ssfr_diskbulge} is that most bulges are growing faster
than their associated outer disk, and this is common for both
pseudobulges {\em and} classical bulges.  In Fig.~\ref{fig:drivers} we
replot Fig.~\ref{fig:ssfr_diskbulge} which compares the specific SFR
of the bulge to that of the disk, however this time the symbols
represent the type of disk in the galaxy (barred galaxies are
represented by magenta triangles, ovaled galaxies are represented by
cyan circles and galaxies with neither bars nor ovals are represented
by black squares). The solid lines indicates the line of equality; the
dashed line indicates bulge growth that is $10\times$ the growth of
the disk. For all types of galaxy disks (barred, ovaled, and disks
with neither bars nor ovals) the bulge is growing faster than the
disk. Further, the growth is not signifigantly more pronounced in
barred and ovaled galaxies. This does not mean that bars are not
important. Although the exact conditions for bar dissolution are not
well understood, we know that increasing $B/T$ can cause a bar to fade
\citep{shen2004,athan2005b}. It may be that many galaxies that are
driving faster growth in $B/T$ destroy their bars faster.  {\em
  Nonetheless, we observe that most galaxies in our sample are
  increasing $B/T$ through present day star formation. If this
  evolution is due to internal rearrangement of disk gas and stars
  (i.e.~secular evolution), this implies that secular evolution is a
  universal process, occurring in every giant galactic disk.}
\cite{kormendyfisher2005} discuss the universality of secular
evolution. They argue that the processes that drive the internal
growth of bulges arise from natural tendencies of self gravitating
disks. If secular evolution is what is driving bulge growth in our
sample, then it comes as no suprise that it appears common in
intermediate-type galaxies.  \cite{droryfisher2007} show that if a
galaxy contains a classical bulge, the entire galaxy is on the red
sequence. If secular evolution is occurring in all galaxies with
disks, those galaxies who no longer have as much fuel for significant
star formation, namely red sequence galaxies, would simply not grow as
much.



The time scales of pseudobulge growth we observe are similar to the
time-scale of bulge formation in simulations
\citep{debattista2004,heller2007a,heller2007b}.  Yet present-day star
formation in more massive pseudobulges in our sample can only account
for half of their stellar mass within a reasonable time-frame. It is
likely that the SFR of pseudobulges were higher in the past just as
higher historic SFRs in the past are typical in disk galaxies
\citep{kennicutt1994}. Also, simulations show that there is a
signifigant amount of radial transfer of stellar mass that occurs
naturally within disk galaxies \cite{roskar2008}. Taking all of these
different factors into account, it seems quite reasonable to us to
assume that the present-day SFR need only accound for some fraction,
possibly of the order of one-half, of the mass of the bulge. {\em
  Thus, the star formation rates we observe in nearby pseudobulges are
  sufficient to form their stellar mass, and are not so high as to
  require some mechanism to shut off secular evolution in many
  galaxies.}

Also, it is quite possible that the stellar mass in present day
pseudobulges arises from multiple evolutionary mechanisms.
\cite{carollo2007} find an underlying population of old stars in many
very late-type bulges.  \cite{cox2008} show that very minor mergers do
not signifigantly alter the SFR of disk galaxies, thus it may be that
part of the mass is directly accreted while or before a pseudobulge is
built in the same galaxy. Also, clump instabilities occur frequently
in simulations \citep[e.g.][]{noguchi1999,immeli2004,debattista2006},
and the clumps fall to the center of the galaxy. The result is a
central density that is higher than the inward extrapolation of the
disk profile.  These clumps may have been observed in high redshift
galaxies \citep{bournard2008}. However, \cite{elmegreen2008} show that
these clumps genarally heat the disk, and produce structures looking
more similar to classical bulges than to pseudobulges. Also, accretion
of mass to make half the bulge mass is likely to heat a disk
\citep{toth1992,velazquez1999}, therefore stabilizing the disk against
efficient mechanisms to drive more radial gas and mass inflow.


\begin{figure}[t!]
  \centering
\includegraphics[width=.5\textwidth]{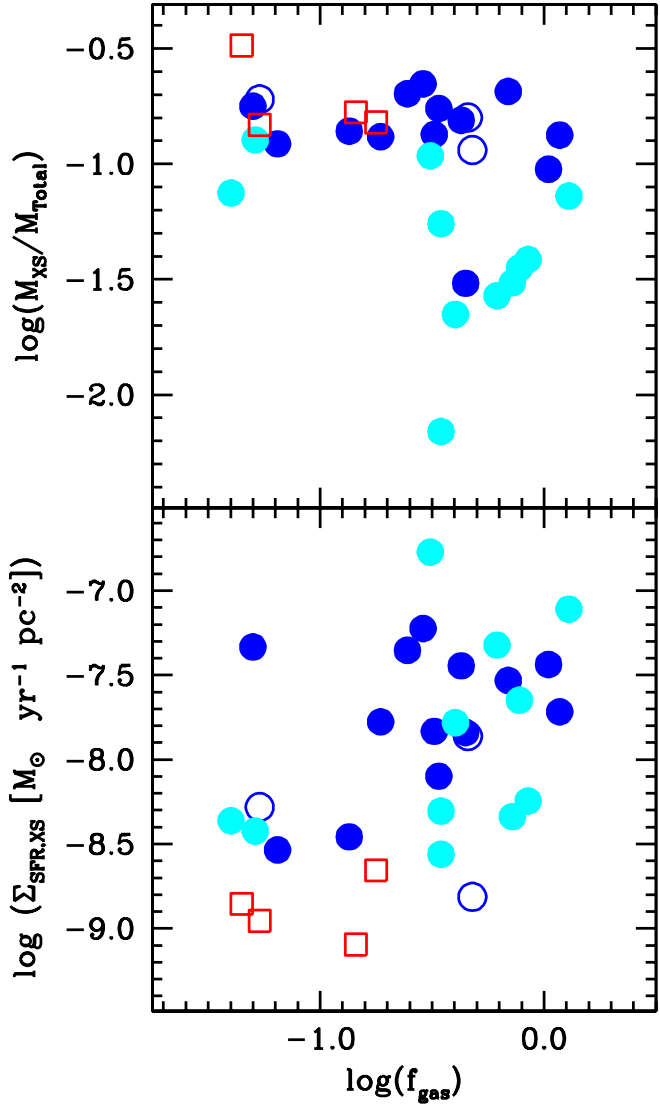}
\caption{Here we show the bulge-to-total ratio of the stellar mass and
  surface desnity of SFR to the fraction of the gas fraction
  ($M_{gas}/M_{stars}$).  The black line represents the line of
  equality. Symbols are the same as
  Fig.~\ref{fig:sfr_xs}. \label{fig:fgas}}
\end{figure}

\subsection{Future Pseudobulge Growth In Late-Type Galaxies}

If secular evolution is responsible for pseudobulges, then it makes
sense that there is a distribution of bulge-to-total ratios that
extends all the way to zero. \cite{spitzer1} show that late-type
bulges and pseudobulges form a sequence in the mass versus surface
density plane; this is reproduced in Fig.~\ref{fig:struct}. Also,
late-type bulges are roughly the same size as pseudobulges (as
measured by $r_{XS}$).  Thus, it appears that adding stellar mass to
late-type bulges would make them similar to pseudobulges in
intermediate-type galaxies.

Is there enough gas in the bulges of late-type galaxies to build a
pseudobulge in the future? We combine the central surface density of
gas from BIMA SONG \citep{helfer2003,sheth2005} and our SFR densities
to determine gas consumption time scales. Given the small sample this
produces, the results should only be taken as suggestive. A more
rigorous study is needed for a more accurate analysis.  We find gas
consumption times of 5-8~Gyr, very similar to the typical pseudobulge
doubling time which is about 5~Gyr. However, some of the smaller
late-type bulges are an order of magnitude smaller than the typical
pseudobulge.

We combine our sample with nuclear gas masses from \cite{sheth2005}; we find
that if late-type pseudobulges continue consuming the gas in their
centers at the same rates as today, only the few with the highest
nuclear gas masses (which typically have $\psi_{XS}\sim
0.2$~M$_{\odot}$~yr$^{-1}$) will be able to build a pseudobulge before
running out of gas in the center. Thus if the smaller pseudobulges in
late-type galaxies with lower SFR are to build larger pseudobulges,
then gas must be driven inward from the outer disk. However, as
discussed by \cite{sheth2005} it is not trivial to get the gas to the
center of the galaxy.  Nonetheless, we ask if there is a large enough
reservoir of gas in the whole galaxy to supply late-type bulges with
enough gas to grow a pseudobulges more similar to those in
intermediate-type galaxies.

Some late-type bulges have capacity to grow into typical pseudobulges,
but not all.  In Fig.~\ref{fig:fgas}, we compare the gas mass
fraction, $f_{gas}\equiv M_{gas}/M_{stars}$, to bulge-to-total ratio
(top panel) and the SFR density of the bulge (bottom panel). Note that
the quantity $f_{gas}$ is describing the entire galaxy, not just the
bulge.  We use gas masses reported in \citep{sheth2005} and
\citep{kennicutt98}.  In this sample, late-type pseudobulges have
similar total gas fractions on average as the population of
pseudobulges. There is enough gas in the entire galaxy in the lowest
$B/T$ pseudobulges systems to build a bulge with $B/T\sim0.1$ (given
that the process of gas inflow and conversion into stars can maintain
an efficiency of order 10\%). However, at the current SFR this would
take longer than a Hubble time for smaller late-type pseudobulges.

There is some evidence that disk mass plays an important role. In
Fig.~\ref{fig:diskmass} we show that massive bulges exist only in
massive disks, and that the spread in $B/T$ increases for lower-mass
disks. Yet, M~101 is an unbarred Sc galaxy with total stellar mass
$M\sim3\times10^9~M_{\odot}$, a total SFR of $\psi_{total}\sim
0.4~M_{\odot}~yr^{-1}$ and gas fraction of $f_{gas}\sim 0.5$. These
three properties are similar to M~63, an unbarred Sbc galaxy with
$M\sim8\times10^9~M_{\odot}$, a total SFR of $\psi_{total}\sim
0.6~M_{\odot}~yr^{-1}$, and gas fraction of $f_{gas}\sim 0.6$. The
total mass is less than a factor of 2 different, the total SFR is
similar, yet M~63 has a pseudobulge mass of $M_{XS}\sim10^9~M_{\odot}$
but M~101 has a bulge that is 2 orders of magnitude smaller,
$M_{XS}\sim 2\times10^{7}~M_{\odot}$. It seems that having a large
disk mass is necessary, but not sufficient to form a large
pseudobulge.

Bars in early type galaxies are longer than bars in late-type galaxies
\citep{erwin2005}. It may be that bars observed in Sa-Sbc galaxies are
able to fuel more active growth of pseudobulges than bars in later
types.  However, \cite{combes1993} show with simulations that as bars
grow they also slow down and become longer; therefore it seems
possible that the difference in bar types is a consequence of the
difference in bulge mass, or at least they arise from a common
process. Also recall from Fig.~\ref{fig:drivers}, that barred galaxies
are not showing faster $B/T$ growth. It is possible that some other
process such as external accretion of satellites could supply the disk
with cold gas that may foster secular evoution
\citep[see][]{bournaud2002}. It may be worth noting that warps in
disks due to accretion survive in simulations for times that are
comparable to the typical growth time of a pseuodbulges, a few Gyr
\citep{shen2006}. If accretion triggers internal evolution, this
generates a seeemingly arbitrary, and possibly unobservable,
distinction between those galaxies that form large pseudobulges
($B/T\sim 15$\%) and those with almost no bulge at all, as in M~101.

Other possibilities for generating the differences between galaxies
with massive pseudobulges and those with small pseudobulges include
dark matter halo-triaxiality or other couplings between baryonic
mass and dark matter properties \citep[e.g.]{foyle2008}.  Finally, it may
just be a matter of time. Fig.~\ref{fig:diskmass} shows that late-type
bulges and pseudobulges at the same disk mass have the same SFR
density; if internal evolution continues driving gas to the centers of
late-type disk galaxies it may be that in a few Gyr M~101 will look
similar to galaxies like NGC~5055 in $B/T$ as well as in other
properties.

\subsection{Inactive Pseudobulges or Acitve Classical Bulges?}

In \S\ref{sec:sfr} we distinguish galaxies based upon two seperate
properties. First, galaxies are seperated via morphology; we call
those bulges that possess disk-like structure, as outlined in
\cite{kk04}, pseudobulges and those with bulges that better
resemble E-type galaxies classical bulges. However, we notice that
the set of pseudobulges has two subsets: those that actively form
stars and those that are inactive.

In all cases involving star formation rates, inactive pseudobulges are
found between (active) pseudobulges and classical bulges. Yet, their
nuclear morphology is similar to pseudobulges. Furthermore, they have
S\'ersic index less than two, which strengthens the claim that they
are pseudobulges \citep{fisherdrory2008}. However, in the structural
parameter correlations, presented in \cite{spitzer1}, when we
distinguish the bulges based not just on morphology but also specific
SFR the inactive pseudobulges appear more like classical bulges than
pseudobulges. Their true nature is thus, somewhat uncertain.

Inactive pseudobulges seem to be transition objects in parameter
space, and possibly in formation mechanism. In
Fig.~\ref{fig:ssfr_diskbulge} we show that all bulges in our sample
are growing faster than their outer disk, and this is independent
of the type of bulge. Thus, if secular evolution is driving this
trend, and if the bulge is small enough and the disk has enough
gas, it is possible that a pseudobulge could grow on top of the
classical bulge. In this case the bulge mass would be high with
respect to the SFR, because a large fraction of the mass is in a
classical bulge. This argument is supported by the fact that inactive
pseudobulges typically have higher $B/T$ than active pseudobulges.

Secondly, it is also possible that inactive pseudobulges are galaxies
in which secular evolution is effectively shutting off. We note again
that inactive pseudobulges are the largest $B/T$ pseudobulges, thus it
is possible that the disk has built a large bulge that now stabilzes the
disk against large scale instabilities, as has been seen in many
simulations \citep[e.g.][]{friedli1993,shen2004,athan2005b}.  However,
this would not explain why inactive pseudobulges look more similar to
classical bulges in structural parameter correlations.

It is possible that we are seeing externally driven star formation in
a few classical bulges. This would explain why the structural
properties of inactive pseudobulges are so similar to classical
bulges, seen in Fig.~\ref{fig:struct}. Also, it appears that in both
the $\psi_{XS}-M_{XS}$ and $\psi_{XS}/M_{XS}-M_{XS}$ planes
(Fig.~\ref{fig:sfr_xs}) inactive pseudobulges show similar behavior to
classical bulges that is shifted slightly toward higher SFR.  We note
that M~81, which is denoted in each figure as a red triangle, has a
classical bulge and the galaxy is known to be interacting with nearby
M~82.  Thus, it has dust in the bulge that is easily seen in MIPS and
IRAC~8~$\mu$m images \citep{gordon2004}. Inactive pseudobulges have
different optical morphology and also much higher specific SFR than
M~81. However, since M~81 is the largest bulge in our sample, it is
probably not the best comparison object. This would imply that the
method classifying bulges based on the presence of disk-like morpholgy
may be flawed in this respect, and that 6 out of 22 pseudobulges in
our sample would be misclassified. However, it would be hard to
explain why inactive pseudobulges have S\'ersic indices below 2 just
like pseudobulges and unlike any classical bulges.

Given these three possibilities we do not know what the true nature of
inactive pseuodbulges is. It is quite possible that inactive
pseudobulges are a mixed bag of objects, that some are evolved
pseudobulges, others true composites, and some are active classical
bulges. Furture work involving dynamics may be more revealing of their
physical nature. Nonetheless, the existance of inactive pseuodbulges
in no way denies the fact that active pseuodbulges are growing
rapidly. We note in Tables~1~\&~2 that these galaxies do not have
significantly perturbed global morphology.  Recall that in
Fig.~\ref{fig:sfr_xs} we show that some pseudobulges have sufficient
SFR to double their stellar mass in 1-2~Gyr. If this star formation
were due to non-secular means, (namely mergers) it is unlikely that the
merger remnant would have relaxed so much as to form a cold disk with
a central bulge yet star burst. 


\acknowledgments

DBF wishes to thank Prof.\ J.~Kormendy and the University of Texas at
Austin as well as Prof.\ A.~Filippenko and the University of
California at Berkeley for providing support. ND, MHF and DBF thank
the Max-Planck Society for support during this project. We also thank
J.~Kormendy, V.~Debattista, R.~Bender and P.~Erwin for stimulating and
valuable discussions. We thank the referee for his/her helpful
comments that greatly imporved the quality of this work.

This work is based on observations made with the Spitzer Space
Telescope, which is operated by the Jet Propulsion Laboratory,
California Institute of Technology under a contract with NASA. Support
for this work was provided by NASA through an award issued by
JPL/Caltech. DBF acknowledges support by the National Science
Foundation under grant AST 06-07490.

Based on observations made with the NASA/ESA Hubble Space Telescope,
obtained at the Space Telescope Science Institute, which is operated
by the Association of Universities for Research in Astronomy, Inc.,
under NASA contract NAS 5-26555.  Some of the data presented in this
paper were obtained from the Multi-mission Archive at the Space
Telescope Science Institute (MAST). STScI is operated by the
Association of Universities for Research in Astronomy, Inc., under
NASA contract NAS5-26555. Support for MAST for non-HST data is
provided by the NASA Office of Space Science via grant NAG5-7584 and
by other grants and contracts.

This research has made use of the NASA/IPAC Extragalactic Database
(NED) which is operated by the Jet Propulsion Laboratory, California
Institute of Technology, under contract with the National Aeronautics
and Space Administration.

This publication makes use of data products from the Two Micron All
Sky Survey, which is a joint project of the University of
Massachusetts and the Infrared Processing and Analysis
Center/California Institute of Technology, funded by the National
Aeronautics and Space Administration and the National Science
Foundation.



\begin{deluxetable}{lccclcccccc}

  \tablewidth{0pt} \tablecaption{Sample Galaxy Properties}
  \tablehead{\colhead{Identifier} & \colhead{Alt.} & \colhead{Dist.} &
    \colhead{Bulge}
    & \colhead{Hubble} &\colhead{Disk} & \colhead{M$_B$} & \colhead{$n_{b}(V)$\tablenotemark{d}}& \colhead{$\psi_{Total}$} & \colhead{Total Stellar Mass} & \colhead{$M_{gas}/M_{star}$\tablenotemark{e}}\\
    \colhead{} & \colhead{Name} & \colhead{(Mpc)} &
    \colhead{Type\tablenotemark{a}} & \colhead{Type\tablenotemark{b}}
    &\colhead{Type\tablenotemark{c}} & \colhead{(B mags)} & \colhead{ } &
    \colhead{(M$_{\odot}$~yr$^{-1}$)} &
    \colhead{($10^9$~M$_{\odot}$)} & \colhead{ } } \startdata
NGC~1617 &  & 13.78 & C & Sa & B & -19.5 & 2.04 $\pm$ 0.22 $^{\dagger}$ & 0.03 $\pm$ 0.002 & 12.7 $\pm$ 0.9 & ...  \\
NGC~2775 &  & 14.42 & C & SB0/a & U & -19.7 & 3.80 $\pm$ 0.19  & 0.09 $\pm$ 0.013 & 16.5 $\pm$ 1.1 & ...  \\
NGC~2841 &  & 8.96 & C & Sb & U & -19.7 & 2.15 $\pm$ 0.24  & 0.23 $\pm$ 0.027 & 14.1 $\pm$ 0.5 & 0.29 (1) \\
NGC~3031 & M~81 & 3.91 & C & Sb & U & -20.2 & 3.79 $\pm$ 0.20  & 0.36 $\pm$ 0.020 & 22.6 $\pm$ 2.3 & 0.07 (1) \\
NGC~4450 &  & 14.28 & C & Sab & B & -19.9 & 3.67 $\pm$ 0.29 $^{\dagger}$ & 0.07 $\pm$ 0.007 & 31.8 $\pm$ 1.8 & 0.09 (1) \\
NGC~4698 &  & 15 & C & Sa & B & -19.3 & 3.60 $\pm$ 0.26  & 0.03 $\pm$ 0.004 & 56.4 $\pm$ 2.7 & 0.01 (2) \\
NGC~4725 &  & 16.32 & C & Sa & B & -21.1 & 5.23 $\pm$ 0.18  & 0.21 $\pm$ 0.006 & 31.5 $\pm$ 1.1 & 0.24 (1) \\
NGC~6744 &  & 10.28 & C & SBbc & B & -20.9 & 2.53 $\pm$ 0.30  & 0.27 $\pm$ 0.021 & 12.3 $\pm$ 0.5 & ...  \\
NGC~3368 & M~96 & 13 & P(I) & Sab & O & -20.5 & 1.71 $\pm$ 0.37  & 0.20 $\pm$ 0.011 & 21.5 $\pm$ 2.2 & 0.09 (1) \\
NGC~3953 &  & 13.24 & P(I) & SBbc & B & -20.1 & 2.69 $\pm$ 0.37 $^{\dagger}$ & 0.15 $\pm$ 0.014 & 11.2 $\pm$ 1.2 & 0.77 (1) \\
NGC~4274 &  & 13.17 & P(I) & Sa & O & -19.3 & 1.82 $\pm$ 0.24  & 0.14 $\pm$ 0.002 & 13.7 $\pm$ 0.4 & ...  \\
NGC~7177 &  & 16 & P(I) & Sab & B & -19.2 & 1.51 $\pm$ 0.18  & 0.06 $\pm$ 0.003 & 6.3 $\pm$ 0.5 & ...  \\
NGC~7217 &  & 16.63 & P(I) & Sb & U & -20.1 & 3.52 $\pm$ 0.20  & 0.20 $\pm$ 0.010 & 23.3 $\pm$ 1.3 & ...  \\
NGC~7331 &  & 13.24 & P(I) & Sb & U & -20.4 & 4.53 $\pm$ 0.45 $^{\dagger}$ & 2.09 $\pm$ 0.322 & 31.1 $\pm$ 3.3 & 0.74 (1) \\
NGC~1433 &  & 9.85 & P & SBb & B & -19.2 & 0.90 $\pm$ 0.13 $^{\dagger}$ & 0.06 $\pm$ 0.002 & 3.8 $\pm$ 0.2 & ...  \\
NGC~1512 &  & 9.85 & P & SBb & B & -18.8 & 1.56 $\pm$ 0.14  & 0.20 $\pm$ 0.002 & 4.0 $\pm$ 0.2 & ...  \\
NGC~1672 &  & 12.3 & P & Sb & O & -20.1 & 1.24 $\pm$ 0.11 $^{\dagger}$ & 1.39 $\pm$ 0.019 & 7.2 $\pm$ 0.3 & ...  \\
NGC~3351 & M~95 & 7.06 & P & SBb & B & -18.7 & 1.51 $\pm$ 0.39  & 0.19 $\pm$ 0.001 & 4.3 $\pm$ 0.6 & 0.40 (1) \\
NGC~3521 &  & 7.2 & P & Sbc & U & -19.4 & 3.20 $\pm$ 0.46  & 0.53 $\pm$ 0.039 & 11.8 $\pm$ 1.4 & 0.52 (1) \\
NGC~3593 &  & 9 & P & Sa & U & -17.9 & 1.80 $\pm$ 0.37  & 0.21 $\pm$ 0.001 & 4.4 $\pm$ 0.7 & ...  \\
NGC~3627 & M~66 & 6.83 & P & Sb & B & -19.4 & 2.90 $\pm$ 0.42  & 0.32 $\pm$ 0.015 & 7.2 $\pm$ 0.8 & 1.93 (1) \\
NGC~3675 &  & 10.12 & P & Sb & U & -19.1 & 3.16 $\pm$ 0.29 $^{\dagger}$ & 0.18 $\pm$ 0.026 & 8.5 $\pm$ 0.8 & 0.31 (2) \\
NGC~3726 &  & 13.24 & P & Sbc & B & -19.9 & 1.94 $\pm$ 0.33 $^{\dagger}$ & 0.21 $\pm$ 0.010 & 3.0 $\pm$ 0.5 & 0.73 (2) \\
NGC~4245 &  & 13 & P & SBa & B & -18.2 & 1.90 $\pm$ 0.34  & 0.02 $\pm$ 0.001 & 3.6 $\pm$ 0.3 & ...  \\
NGC~4314 &  & 14 & P & SBa & B & -19.3 & 2.37 $\pm$ 0.39  & 0.12 $\pm$ 0.010 & 6.4 $\pm$ 0.9 & ...  \\
NGC~4380 &  & 13 & P & Sab & U & -17.9 & 1.41 $\pm$ 0.20  & 0.03 $\pm$ 0.008 & 2.8 $\pm$ 0.5 & ...  \\
NGC~4394 &  & 14.28 & P & SBb & B & -19.2 & 1.65 $\pm$ 0.25  & 0.05 $\pm$ 0.004 & 4.3 $\pm$ 0.4 & 0.22 (2) \\
NGC~4448 &  & 10 & P & Sa & B & -18.0 & 1.68 $\pm$ 0.34  & 0.04 $\pm$ 0.001 & 3.3 $\pm$ 0.3 & ...  \\
NGC~4457 &  & 10.22 & P & RSb & O & -18.3 & 1.66 $\pm$ 0.50 $^{\dagger}$ & 0.06 $\pm$ 0.003 & 3.8 $\pm$ 0.2 & ...  \\
NGC~4569 & M~90 & 14.28 & P & Sab & B & -20.7 & 1.90 $\pm$ 0.28  & 0.53 $\pm$ 0.029 & 14.9 $\pm$ 1.6 & 0.69 (1) \\
NGC~4639 &  & 14.28 & P & SBb & B & -18.6 & 1.64 $\pm$ 0.45  & 0.03 $\pm$ 0.004 & 2.3 $\pm$ 0.2 & 0.11 (2) \\
NGC~4736 & M~94 & 4 & P & RSab & O & -19.3 & 1.62 $\pm$ 0.26  & 0.31 $\pm$ 0.015 & 5.8 $\pm$ 0.2 & 0.47 (1) \\
NGC~4826 & M~64 & 7.48 & P & Sab & U & -20.1 & 3.94 $\pm$ 0.34  & 0.27 $\pm$ 0.006 & 16.2 $\pm$ 1.0 & 0.08 (1) \\
NGC~5055 & M~63 & 7.27 & P & Sbc & U & -20.0 & 1.84 $\pm$ 0.24  & 0.62 $\pm$ 0.067 & 12.8 $\pm$ 0.8 & 0.56 (1) \\
NGC~5194 & M~51 & 6.52 & P & Sbc & U & -20.3 & 0.55 $\pm$ 0.07  & 1.46 $\pm$ 0.188 & 9.3 $\pm$ 0.7 & 1.70 (1) \\
NGC~5248 &  & 16.75 & P & Sbc & O & -20.2 & 1.62 $\pm$ 0.27  & 0.92 $\pm$ 0.011 & 13.2 $\pm$ 0.8 & 1.13 (1) \\
NGC~5879 &  & 13.45 & P & Sb & U & -18.5 & 1.65 $\pm$ 0.19  & 0.09 $\pm$ 0.004 & 2.1 $\pm$ 0.1 & ...  \\
IC~342 &  & 2.58 & P(L) & Scd & U & -17.4 & 1.88 $\pm$ 0.41 $^{\dagger}$ & 0.31 $\pm$ 0.019 & 3.1 $\pm$ 0.3 & 0.76 (2) \\
NGC~0628 & M~74 & 9.05 & P(L) & Sc & U & -20.1 & 1.45 $\pm$ 0.10  & 0.54 $\pm$ 0.043 & 4.9 $\pm$ 0.7 & 0.56 (1) \\
NGC~0925 &  & 9.08 & P(L) & SBc & B & -19.2 & 1.18 $\pm$ 0.21 $^{\dagger}$ & 0.45 $\pm$ 0.024 & 1.4 $\pm$ 0.4 & 1.39 (1) \\
NGC~2403 &  & 3.35 & P(L) & Sc & U & -18.8 & 1.50 $\pm$ 0.62 $^{\dagger}$ & 0.30 $\pm$ 0.118 & 1.3 $\pm$ 0.2 & 0.08 (2) \\
NGC~2903 &  & 8.16 & P(L) & Sc & U & -20.0 & 0.42 $\pm$ 0.07  & 0.76 $\pm$ 0.018 & 7.1 $\pm$ 0.6 & 0.51 (1) \\
NGC~3184 &  & 9.11 & P(L) & Sc & U & -19.4 & 1.78 $\pm$ 0.48 $^{\dagger}$ & 0.07 $\pm$ 0.003 & 3.3 $\pm$ 0.2 & 0.64 (1) \\
NGC~3198 &  & 9.11 & P(L) & Sc & U & -18.9 & 1.69 $\pm$ 0.63 $^{\dagger}$ & 0.39 $\pm$ 0.009 & 1.7 $\pm$ 0.1 & ...  \\
NGC~3769 &  & 14.04 & P(L) & SBc & B & -18.2 & 0.54 $\pm$ 0.09 $^{\dagger}$ & 0.14 $\pm$ 0.010 & 1.3 $\pm$ 0.8 & ...  \\
NGC~3938 &  & 14.04 & P(L) & Sc & B & -19.8 & 1.68 $\pm$ 0.34 $^{\dagger}$ & 0.30 $\pm$ 0.044 & 4.6 $\pm$ 2.6 & 1.16 (1) \\
NGC~4136 &  & 12.48 & P(L) & Sc & B & -18.4 & 0.58 $\pm$ 0.65 $^{\dagger}$ & 0.10 $\pm$ 0.025 & 0.9 $\pm$ 0.6 & ...  \\
NGC~4303 & M~61 & 19.77 & P(L) & Sc & B & -21.3 & 0.96 $\pm$ 0.14 $^{\dagger}$ & 1.99 $\pm$ 0.092 & 26.2 $\pm$ 2.9 & ...  \\
NGC~4321 & M~100 & 14.28 & P(L) & Sc & B & -20.8 & 0.50 $\pm$ 0.06 $^{\dagger}$ & 1.20 $\pm$ 0.024 & 16.4 $\pm$ 6.2 & ...  \\
NGC~4414 &  & 12.48 & P(L) & Sc & U & -19.5 & 2.79 $\pm$ 0.31 $^{\dagger}$ & 0.49 $\pm$ 0.211 & 19.7 $\pm$ 3.3 & 1.24  \\
NGC~4559 &  & 9.87 & P(L) & Sc & U & -19.7 & 1.85 $\pm$ 0.82 $^{\dagger}$ & 0.61 $\pm$ 0.398 & 10.0 $\pm$ 0.1 & 0.06 (1) \\
NGC~4580 &  & 14.63 & P(L) & Sc/Sa & U & -18.1 & 1.65 $\pm$ 0.60 $^{\dagger}$ & 0.05 $\pm$ 0.001 & 2.4 $\pm$ 1.0 & ...  \\
NGC~5457 & M~101 & 5.03 & P(L) & Sc & U & -20.1 & 1.83 $\pm$ 0.46 $^{\dagger}$ & 0.44 $\pm$ 0.073 & 4.2 $\pm$ 0.2 & 0.57 (1) \\
NGC~6946 &  & 5.53 & P(L) & Sc & U & -19.0 & 1.87 $\pm$ 0.36  & 0.58 $\pm$ 0.042 & 5.8 $\pm$ 1.5 & 2.11 (1) \\
 \enddata
 \tablenotetext{a}{C -- classical bulge; P -- pseudobulge; P(I) -- pseudobulge designated as inactive; P(L) -- pseudobulge in a late-type galaxy.}
 \tablenotetext{b}{Taken from Sandage \& Bedke (1994)}
 \tablenotetext{c}{B -- Barred Disk; O -- Ovalled Disk; U -- Unbarred \& Onovalled.}
 \tablenotetext{d}{$^{\dagger}$ indicates new decomposition; otherwise $n_b$ is taken from \cite{fisherdrory2008}.}
 \tablenotetext{e}{Sources are as follows: (1)--\cite{sheth2005}; (2)--\cite{kennicutt98}}
 \end{deluxetable}

\clearpage
\begin{landscape}
\LongTables
\begin{deluxetable}{lccccccccccc}
\tabletypesize{\footnotesize}
  \tablewidth{0pt} \tablecaption{Bulge Properties}
  \tablehead{\colhead{Identifier} & \colhead{Alt.} & \colhead{Bulge} &
    \colhead{$R_{XS}$} & \colhead{L$_{XS}$(3.6~$\mu$m)} &
    \colhead{L$_{XS}$(24~$\mu$m)} &
    \colhead{L$_{XS}$(FUV)} & \colhead{$\psi_{XS}$} & \colhead{$\Sigma_{SFR,XS}$} & \colhead{$M_{XS}$} & \colhead{$M_{Disk}$}  & \colhead{$\Sigma_{XS}$} \\
    \colhead{ } & \colhead{Name} & \colhead{Type\tablenotemark{a}} &
    \colhead{(kpc)} & \colhead{(10$^{40}$~erg~s$^{-1}$)} &
    \colhead{(10$^{40}$~erg~s$^{-1}$)} &
    \colhead{(10$^{40}$~erg~s$^{-1}$)} & \colhead{(10$^{-3}$ M$_{\odot}$ yr$^{-1}$)}  
   & \colhead{( M$_{\odot}$ yr$^{-1}$ pc$^{-2}$)} & \colhead{($10^8$ M$_{\odot}$)}& \colhead{($10^8$ M$_{\odot}$)} & \colhead{(M$_{\odot}$ pc$^{-2}$)} } \startdata
NGC~6744 &  & C & 2.5 $\pm$ 0.21 & 1.4 $\pm$ 0.14 & 1.8 $\pm$ 0.1 & ...    & 8.1 $\pm$ 0.6 & 8.1 $\pm$ 0.6 & 22.79 $\pm$ 1.71 & 75.4 $\pm$ 5.6 & 57.12 $\pm$ 4.68 \\
NGC~3031 & M~81 & C & 4.07 $\pm$ 0.21 & 9.15 $\pm$ 0.37 & 15.9 $\pm$ 0.6 & 7.50 $\pm$ 1.19 & 39.0 $\pm$ 5.5 & 39.0 $\pm$ 5.5 & 38.85 $\pm$ 2.58 & 96.8 $\pm$ 6.4 & 10.35 $\pm$ 0.75 \\
NGC~4450 &  & C & 2.86 $\pm$ 0.16 & 6.11 $\pm$ 0.35 & 8.8 $\pm$ 0.7 & 6.44  0.81 & 23.0 $\pm$ 2.7 & 23.0 $\pm$ 2.7 & 21.38 $\pm$ 0.71 & 90.4 $\pm$ 3.0 & 20.65 $\pm$ 1.39 \\
NGC~2775 &  & C & 3.46 $\pm$ 0.22 & 4.82 $\pm$ 0.32 & 10.2 $\pm$ 1.3 & 17.23 $\pm$ 1.24 & 73.3 $\pm$ 4.1 & 73.3 $\pm$ 4.1 & 72.93 $\pm$ 7.28 & 118.8 $\pm$ 11.9 & 14.01 $\pm$ 1.45 \\
NGC~4725 &  & C & 3.41 $\pm$ 0.21 & 6.97 $\pm$ 0.25 & 12.5 $\pm$ 0.4 & ... $\pm$   & 28.5 $\pm$ 2.9 & 28.5 $\pm$ 2.9 & 46.66 $\pm$ 2.67 & 218.8 $\pm$ 12.5 & 18.16 $\pm$ 1.21 \\
NGC~1617 &  & C & 1.13 $\pm$ 0.07 & 2.68 $\pm$ 0.12 & 2.1 $\pm$ 0.1 & 0.10  0.02 & 5.0 $\pm$ 0.6 & 5.0 $\pm$ 0.6 & 16.48 $\pm$ 0.80 & 421.4 $\pm$ 20.5 & 20.16 $\pm$ 1.51 \\
NGC~2841 &  & C & 1.82 $\pm$ 0.16 & 2.63 $\pm$ 0.26 & 4 $\pm$ 0.4 & 0.45 $\pm$ 0.02 & 28.6 $\pm$ 0.9 & 28.6 $\pm$ 0.9 & 52.97 $\pm$ 1.93 & 212.5 $\pm$ 7.8 & 14.53 $\pm$ 0.71 \\
NGC~4698 &  & C & 1.61 $\pm$ 0.17 & 2.06 $\pm$ 0.17 & 2.2 $\pm$ 0.3 & ... $\pm$   & 7.0 $\pm$ 0.6 & 7.0 $\pm$ 0.6 & 9.70 $\pm$ 0.39 & 100.9 $\pm$ 4.1 & 4.95 $\pm$ 0.38 \\
NGC~3368 & M~96 & P(I) & 1.77 $\pm$ 0.15 & 5 $\pm$ 0.37 & 21.3 $\pm$ 1 & 2.15 $\pm$ 0.25 & 51.9 $\pm$ 2.7 & 51.9 $\pm$ 2.7 & 40.76 $\pm$ 4.09 & 131.9 $\pm$ 13.2 & 41.50 $\pm$ 4.52 \\
NGC~3953 &  & P(I) & 1.83 $\pm$ 0.15 & 1.82 $\pm$ 0.2 & 4.5 $\pm$ 0.3 & ...    & 15.8 $\pm$ 1.5 & 15.8 $\pm$ 1.5 & 12.90 $\pm$ 1.42 & 86.0 $\pm$ 9.5 & 12.27 $\pm$ 1.44 \\
NGC~4274 &  & P(I) & 2.43 $\pm$ 0.05 & 4.67 $\pm$ 0.13 & 26.1 $\pm$ 0.3 & 2.27 $\pm$ 0.06 & 62.8 $\pm$ 0.8 & 62.8 $\pm$ 0.8 & 39.95 $\pm$ 1.09 & 69.6 $\pm$ 1.9 & 21.58 $\pm$ 0.63 \\
NGC~7177 &  & P(I) & 1.36 $\pm$ 0.08 & 2.83 $\pm$ 0.11 & 11.2 $\pm$ 0.4 & 0.93 $\pm$ 0.13 & 26.9 $\pm$ 1.2 & 26.9 $\pm$ 1.2 & 22.45 $\pm$ 1.82 & 31.2 $\pm$ 2.5 & 38.60 $\pm$ 3.41 \\
NGC~7217 &  & P(I) & 4.43 $\pm$ 0.28 & 8.81 $\pm$ 0.44 & 53.6 $\pm$ 2.6 & 3.32 $\pm$ 0.22 & 125.9 $\pm$ 6.2 & 125.9 $\pm$ 6.2 & 68.36 $\pm$ 3.76 & 130.7 $\pm$ 7.2 & 11.08 $\pm$ 0.70 \\
NGC~7331 &  & P(I) & 1.5 $\pm$ 0.1 & 6.96 $\pm$ 0.48 & 31.8 $\pm$ 5.9 & 12.86 $\pm$ 1.00 & 98.8 $\pm$ 15.2 & 98.8 $\pm$ 15.2 & 49.11 $\pm$ 5.17 & 229.0 $\pm$ 24.1 & 69.35 $\pm$ 7.59 \\
NGC~1433 &  & P & 0.59 $\pm$ 0.13 & 1.28 $\pm$ 0.08 & 6.1 $\pm$ 0.2 & ...    & 25.2 $\pm$ 0.9 & 25.2 $\pm$ 0.9 & 9.22 $\pm$ 0.54 & 24.6 $\pm$ 1.5 & 83.44 $\pm$ 8.17 \\
NGC~1512 &  & P & 1.42 $\pm$ 0.07 & 1.18 $\pm$ 0.04 & 9.8 $\pm$ 0.1 & 71.88 $\pm$ 0.76 & 180.4 $\pm$ 2.0 & 180.4 $\pm$ 2.0 & 8.55 $\pm$ 0.43 & 26.4 $\pm$ 1.3 & 13.51 $\pm$ 0.79 \\
NGC~1672 &  & P & 1.18 $\pm$ 0.11 & 3.6 $\pm$ 0.12 & 244.9 $\pm$ 2.4 & 39.90 $\pm$ 1.36 & 629.4 $\pm$ 8.4 & 629.4 $\pm$ 8.4 & 19.35 $\pm$ 0.78 & 60.2 $\pm$ 2.4 & 44.18 $\pm$ 2.90 \\
NGC~3351 & M~95 & P & 0.97 $\pm$ 0.05 & 1.26 $\pm$ 0.04 & 51.2 $\pm$ 0.2 & 6.31 $\pm$ 0.06 & 127.0 $\pm$ 0.6 & 127.0 $\pm$ 0.6 & 8.78 $\pm$ 1.31 & 30.4 $\pm$ 4.6 & 30.01 $\pm$ 4.68 \\
NGC~3521 &  & P & 1.07 $\pm$ 0.06 & 2.16 $\pm$ 0.25 & 21.1 $\pm$ 1.5 & 2.86 $\pm$ 0.24 & 52.9 $\pm$ 3.9 & 52.9 $\pm$ 3.9 & 15.78 $\pm$ 1.83 & 86.7 $\pm$ 10.1 & 43.98 $\pm$ 5.27 \\
NGC~3593 &  & P & 2.3 $\pm$ 0.13 & 2.26 $\pm$ 0.08 & 82.6 $\pm$ 0.3 & ...    & 206.2 $\pm$ 1.2 & 206.2 $\pm$ 1.2 & 19.15 $\pm$ 3.03 & 17.7 $\pm$ 2.8 & 11.57 $\pm$ 1.88 \\
NGC~3627 & M~66 & P & 0.8 $\pm$ 0.1 & 1.48 $\pm$ 0.1 & 17.1 $\pm$ 0.8 & 0.18 $\pm$ 0.03 & 38.2 $\pm$ 1.8 & 38.2 $\pm$ 1.8 & 9.66 $\pm$ 1.03 & 58.9 $\pm$ 6.3 & 48.26 $\pm$ 6.30 \\
NGC~3675 &  & P & 1.03 $\pm$ 0.12 & 1.38 $\pm$ 0.14 & 18.5 $\pm$ 2.1 & ...    & 54.9 $\pm$ 7.9 & 54.9 $\pm$ 7.9 & 11.17 $\pm$ 1.09 & 56.3 $\pm$ 5.5 & 33.21 $\pm$ 3.76 \\
NGC~3726 &  & P & 0.81 $\pm$ 0.06 & 0.22 $\pm$ 0.01 & 7.4 $\pm$ 0.3 & ...    & 29.7 $\pm$ 1.4 & 29.7 $\pm$ 1.4 & 0.90 $\pm$ 0.16 & 42.7 $\pm$ 7.6 & 4.36 $\pm$ 0.78 \\
NGC~4245 &  & P & 1.25 $\pm$ 0.09 & 0.76 $\pm$ 0.06 & 4.5 $\pm$ 0.1 & ...    & 15.7 $\pm$ 0.3 & 15.7 $\pm$ 0.3 & 5.99 $\pm$ 0.46 & 23.9 $\pm$ 1.9 & 12.17 $\pm$ 1.13 \\
NGC~4314 &  & P & 1.31 $\pm$ 0.18 & 2.71 $\pm$ 0.28 & 29.8 $\pm$ 1.3 & 18.97 $\pm$ 2.74 & 107.7 $\pm$ 9.0 & 107.7 $\pm$ 9.0 & 19.58 $\pm$ 2.69 & 37.7 $\pm$ 5.2 & 36.24 $\pm$ 5.67 \\
NGC~4380 &  & P & 1.36 $\pm$ 0.15 & 0.33 $\pm$ 0.03 & 2.9 $\pm$ 0.6 & 1.09 $\pm$ 0.32 & 8.8 $\pm$ 2.0 & 8.8 $\pm$ 2.0 & 2.46 $\pm$ 0.41 & 20.8 $\pm$ 3.5 & 4.21 $\pm$ 0.81 \\
NGC~4394 &  & P & 1.05 $\pm$ 0.26 & 0.81 $\pm$ 0.07 & 2.5 $\pm$ 0.2 & ...    & 11.6 $\pm$ 1.0 & 11.6 $\pm$ 1.0 & 6.05 $\pm$ 0.54 & 30.6 $\pm$ 2.7 & 17.41 $\pm$ 1.85 \\
NGC~4448 &  & P & 0.61 $\pm$ 0.07 & 0.83 $\pm$ 0.07 & 3.6 $\pm$ 0.1 & ...    & 15.7 $\pm$ 0.5 & 15.7 $\pm$ 0.5 & 6.12 $\pm$ 0.49 & 22.8 $\pm$ 1.8 & 52.05 $\pm$ 4.73 \\
NGC~4457 &  & P & 0.62 $\pm$ 0.1 & 1.6 $\pm$ 0.08 & 14.4 $\pm$ 0.5 & ...    & 43.9 $\pm$ 2.2 & 43.9 $\pm$ 2.2 & 11.79 $\pm$ 0.61 & 22.2 $\pm$ 1.1 & 97.81 $\pm$ 7.56 \\
NGC~4569 & M~90 & P & 1.37 $\pm$ 0.25 & 3.68 $\pm$ 0.35 & 85.4 $\pm$ 4.3 & 11.66 $\pm$ 0.89 & 214.6 $\pm$ 11.5 & 214.6 $\pm$ 11.5 & 22.87 $\pm$ 2.54 & 125.0 $\pm$ 13.9 & 38.86 $\pm$ 5.17 \\
NGC~4639 &  & P & 0.66 $\pm$ 0.08 & 0.44 $\pm$ 0.04 & 1.9 $\pm$ 0.2 & 0.07 $\pm$ 0.01 & 4.3 $\pm$ 0.5 & 4.3 $\pm$ 0.5 & 2.80 $\pm$ 0.26 & 19.7 $\pm$ 1.8 & 20.61 $\pm$ 2.34 \\
NGC~4736 & M~94 & P & 0.48 $\pm$ 0.02 & 2.05 $\pm$ 0.08 & 12.8 $\pm$ 0.4 & 7.47 $\pm$ 0.54 & 44.9 $\pm$ 2.2 & 44.9 $\pm$ 2.2 & 13.09 $\pm$ 0.49 & 43.7 $\pm$ 1.6 & 177.26 $\pm$ 8.09 \\
NGC~4826 & M~64 & P & 1 $\pm$ 0.09 & 4.09 $\pm$ 0.15 & 54.2 $\pm$ 0.9 & 3.49 $\pm$ 0.38 & 127.5 $\pm$ 2.8 & 127.5 $\pm$ 2.8 & 28.85 $\pm$ 1.72 & 116.7 $\pm$ 7.0 & 91.20 $\pm$ 6.84 \\
NGC~5055 & M~63 & P & 2.5 $\pm$ 0.18 & 3.4 $\pm$ 0.18 & 58.2 $\pm$ 5.5 & 13.22 $\pm$ 2.24 & 157.9 $\pm$ 17.1 & 157.9 $\pm$ 17.1 & 22.38 $\pm$ 1.38 & 99.0 $\pm$ 6.1 & 11.40 $\pm$ 0.82 \\
NGC~5194 & M~51 & P & 1.19 $\pm$ 0.06 & 1.65 $\pm$ 0.12 & 55.1 $\pm$ 7 & 20.22 $\pm$ 2.70 & 166.4 $\pm$ 21.5 & 166.4 $\pm$ 21.5 & 8.85 $\pm$ 0.69 & 96.5 $\pm$ 7.5 & 19.85 $\pm$ 1.59 \\
NGC~5248 &  & P & 1.79 $\pm$ 0.11 & 4.49 $\pm$ 0.16 & 131.2 $\pm$ 1.5 & 0.99 $\pm$ 0.04 & 292.2 $\pm$ 3.4 & 292.2 $\pm$ 3.4 & 27.17 $\pm$ 1.75 & 106.9 $\pm$ 6.9 & 27.03 $\pm$ 1.94 \\
NGC~5879 &  & P & 2.6 $\pm$ 0.19 & 1.76 $\pm$ 0.04 & 28.9 $\pm$ 1.4 & 0.95 $\pm$ 0.06 & 65.9 $\pm$ 3.1 & 65.9 $\pm$ 3.1 & 10.38 $\pm$ 0.33 & 11.5 $\pm$ 0.4 & 4.88 $\pm$ 0.23 \\
IC~342 &  & P(L) & 1.13 $\pm$ 0.29 & 0.18 $\pm$ 0.02 & 60.4 $\pm$ 3.2 & ...    & 191.3 $\pm$ 11.7 & 191.3 $\pm$ 11.7 & 0.83 $\pm$ 0.07 & 40.2 $\pm$ 3.4 & 2.07 $\pm$ 0.40 \\
NGC~0628 & M~74 & P(L) & 1.27 $\pm$ 0.06 & 0.56 $\pm$ 0.03 & 5.8 $\pm$ 0.4 & 5.51 $\pm$ 0.50 & 25.1 $\pm$ 2.0 & 25.1 $\pm$ 2.0 & 2.70 $\pm$ 0.41 & 59.7 $\pm$ 9.0 & 5.34 $\pm$ 0.83 \\
NGC~0925 &  & P(L) & 2.18 $\pm$ 0.1 & 0.13 $\pm$ 0.01 & 6.7 $\pm$ 0.5 & 31.76 $\pm$ 1.60 & 85.1 $\pm$ 4.6 & 85.1 $\pm$ 4.6 & 0.54 $\pm$ 0.15 & 19.2 $\pm$ 5.4 & 0.36 $\pm$ 0.10 \\
NGC~2403 &  & P(L) & 1.51 $\pm$ 0.33 & 0.37 $\pm$ 0.1 & 6.1 $\pm$ 2.5 & 6.20 $\pm$ 2.38 & 27.2 $\pm$ 10.9 & 27.2 $\pm$ 10.9 & 1.60 $\pm$ 0.22 & 15.5 $\pm$ 2.2 & 2.23 $\pm$ 0.38 \\
NGC~2903 &  & P(L) & 0.68 $\pm$ 0.02 & 1.36 $\pm$ 0.01 & 99.4 $\pm$ 2.2 & 10.44 $\pm$ 0.40 & 242.7 $\pm$ 5.7 & 242.7 $\pm$ 5.7 & 7.68 $\pm$ 0.66 & 68.9 $\pm$ 6.0 & 53.45 $\pm$ 6.63 \\
NGC~3184 &  & P(L) & 0.62 $\pm$ 0.08 & 0.14 $\pm$ 0.02 & 4.7 $\pm$ 0.2 & ...    & 20.1 $\pm$ 0.8 & 20.1 $\pm$ 0.8 & 0.72 $\pm$ 0.04 & 37.0 $\pm$ 1.9 & 5.94 $\pm$ 0.66 \\
NGC~3198 &  & P(L) & 0.95 $\pm$ 0.05 & 0.36 $\pm$ 0.02 & 16.5 $\pm$ 0.3 & 1.15 $\pm$ 0.08 & 39.0 $\pm$ 0.9 & 39.0 $\pm$ 0.9 & 1.84 $\pm$ 0.14 & 18.0 $\pm$ 1.4 & 6.55 $\pm$ 0.56 \\
NGC~3769 &  & P(L) & 1.28 $\pm$ 0.07 & 0.32 $\pm$ 0.02 & 11.7 $\pm$ 1 & 8.38 $\pm$ 0.47 & 44.5 $\pm$ 3.2 & 44.5 $\pm$ 3.2 & 1.65 $\pm$ 0.94 & 13.8 $\pm$ 7.9 & 3.19 $\pm$ 1.82 \\
NGC~3938 &  & P(L) & 0.91 $\pm$ 0.05 & 0.28 $\pm$ 0.02 & 5 $\pm$ 0.8 & 0.19 $\pm$ 0.03 & 11.6 $\pm$ 1.7 & 11.6 $\pm$ 1.7 & 1.40 $\pm$ 0.78 & 55.3 $\pm$ 30.7 & 5.35 $\pm$ 3.02 \\
NGC~4136 &  & P(L) & 0.61 $\pm$ 0.21 & 0.02 $\pm$ 0 & 0.6 $\pm$ 0.2 & 1.02 $\pm$ 0.21 & 3.7 $\pm$ 0.9 & 3.7 $\pm$ 0.9 & 0.08 $\pm$ 0.05 & 10.2 $\pm$ 6.5 & 0.69 $\pm$ 0.44 \\
NGC~4303 & M~61 & P(L) & 1.68 $\pm$ 0.32 & 2.19 $\pm$ 0.19 & 101.5 $\pm$ 4.2 & 17.19 $\pm$ 1.30 & 262.3 $\pm$ 12.2 & 262.3 $\pm$ 12.2 & 22.60 $\pm$ 2.49 & 291.5 $\pm$ 32.1 & 25.45 $\pm$ 2.84 \\
NGC~4321 & M~100 & P(L) & 1.63 $\pm$ 0.12 & 2.52 $\pm$ 0.11 & 101.5 $\pm$ 2.2 & 38.98 $\pm$ 0.63 & 310.4 $\pm$ 6.3 & 310.4 $\pm$ 6.3 & 14.29 $\pm$ 5.41 & 162.1 $\pm$ 61.4 & 17.05 $\pm$ 6.83 \\
NGC~4414 &  & P(L) & 0.36 $\pm$ 0.02 & 0.95 $\pm$ 0.04 & 4.2 $\pm$ 1.8 & 0.02 $\pm$ 0.01 & 9.4 $\pm$ 4.1 & 9.4 $\pm$ 4.1 & 6.90 $\pm$ 1.17 & 160.5 $\pm$ 27.2 & 171.70 $\pm$ 32.80 \\
NGC~4559 &  & P(L) & 2.08 $\pm$ 0.79 & 1.84 $\pm$ 0.73 & 12.8 $\pm$ 7.4 & 13.86 $\pm$ 10.10 & 58.9 $\pm$ 38.8 & 58.9 $\pm$ 38.8 & 7.48 $\pm$ 0.03 & 140.2 $\pm$ 0.6 & 5.49 $\pm$ 0.04 \\
NGC~4580 &  & P(L) & 2.17 $\pm$ 0.05 & 0.03 $\pm$ 0 & 11.2 $\pm$ 0.3 & ...    & 43.0 $\pm$ 0.6 & 43.0 $\pm$ 0.6 & 0.22 $\pm$ 0.09 & 21.9 $\pm$ 8.7 & 0.15 $\pm$ 0.07 \\
NGC~5457 & M~101 & P(L) & 1.4 $\pm$ 0.3 & 0.06 $\pm$ 0.02 & 5.4 $\pm$ 0.8 & 2.30 $\pm$ 0.49 & 17.1 $\pm$ 2.9 & 17.1 $\pm$ 2.9 & 0.30 $\pm$ 0.01 & 56.0 $\pm$ 2.0 & 0.48 $\pm$ 0.02 \\
NGC~6946 &  & P(L) & 0.98 $\pm$ 0.08 & 0.97 $\pm$ 0.08 & 94.7 $\pm$ 4.9 & ...    & 232.6 $\pm$ 16.9 & 232.6 $\pm$ 16.9 & 4.27 $\pm$ 1.07 & 75.5 $\pm$ 19.0 & 14.22 $\pm$ 3.58 \\
 \enddata
 \tablenotetext{(a)}{ C -- classical bulge; P -- pseudobulge; P(I) -- inactive pseudobulge; P(L) -- pseudobulge in a late-type galaxy.}
 \end{deluxetable}
\clearpage
\end{landscape}

\newpage
\appendix
\section{New V-band Photometry and  S\'ersic Fits}
\begin{figure*}[t]
\includegraphics[width=\textwidth]{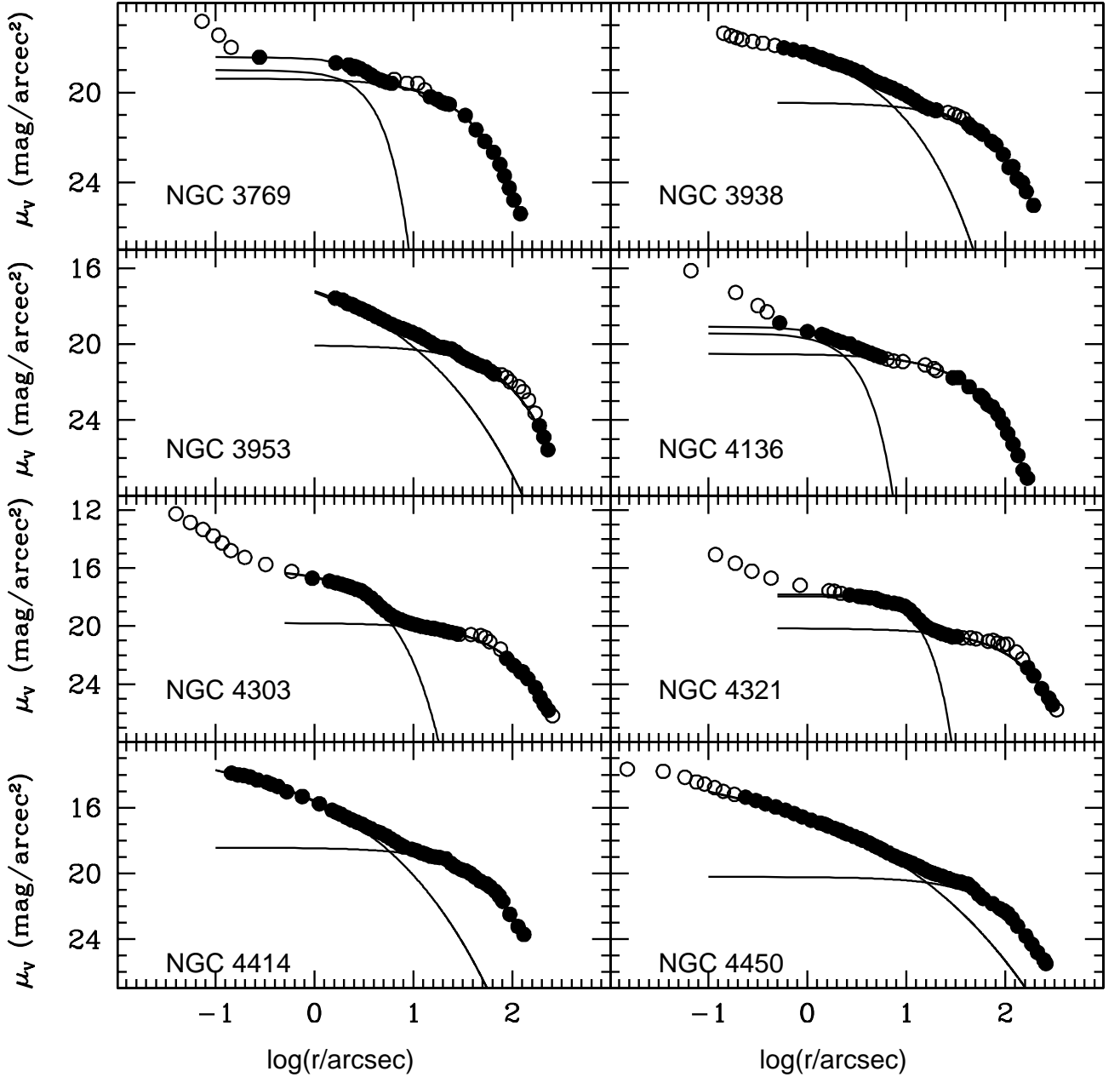}
\caption{Above we show new V-band S\'ersic fits in this paper. Open symbols represent surface brightness isophotes; the filled symbols indicated data elements included in the S\'ersic decomposition. The black lines indicate the fitted function for each galaxy.\label{fig:newprofs}}
\end{figure*}

The method we use to calculate surface brightness profiles and
S\'ersic fits to thos profiles is the same procedure as used in
\cite{fisherdrory2008}. This same procedure is also employed in
\cite{kormendy2008virgo} on elliptical galaxies. Our reduction
software and procedures are discussed in great detail in these two
papers. We refer interested readers to these two papers.

We calculate Johnson $V$-band magnitude zero points using the
transformations in \cite{holtzman1995} for the WFPC2 images and
\cite{sirianni2005} for the ACS images. SDSS g and r profiles are
converted to a single V-band profile for each galaxy using the
transformations in \citet{smithetal2002}. We use colors from
Hyper-LEDA, which refer to colors of the entire galaxies, and the
galaxies in our sample most certainly have non-zero color gradients.
Therefore the absolute values of surface brightness in this paper are
not expected to be consistent to more than 0.3~mag. However, this does
not affect our conclusions which are based the structure in the
profiles and not on absolute magnitude. We check that our magnitudes
are consistent with appature photometry published in the RC3 and
Hyper-LEDA.

\setcounter{figure}{12}
\begin{figure*}
\includegraphics[width=\textwidth]{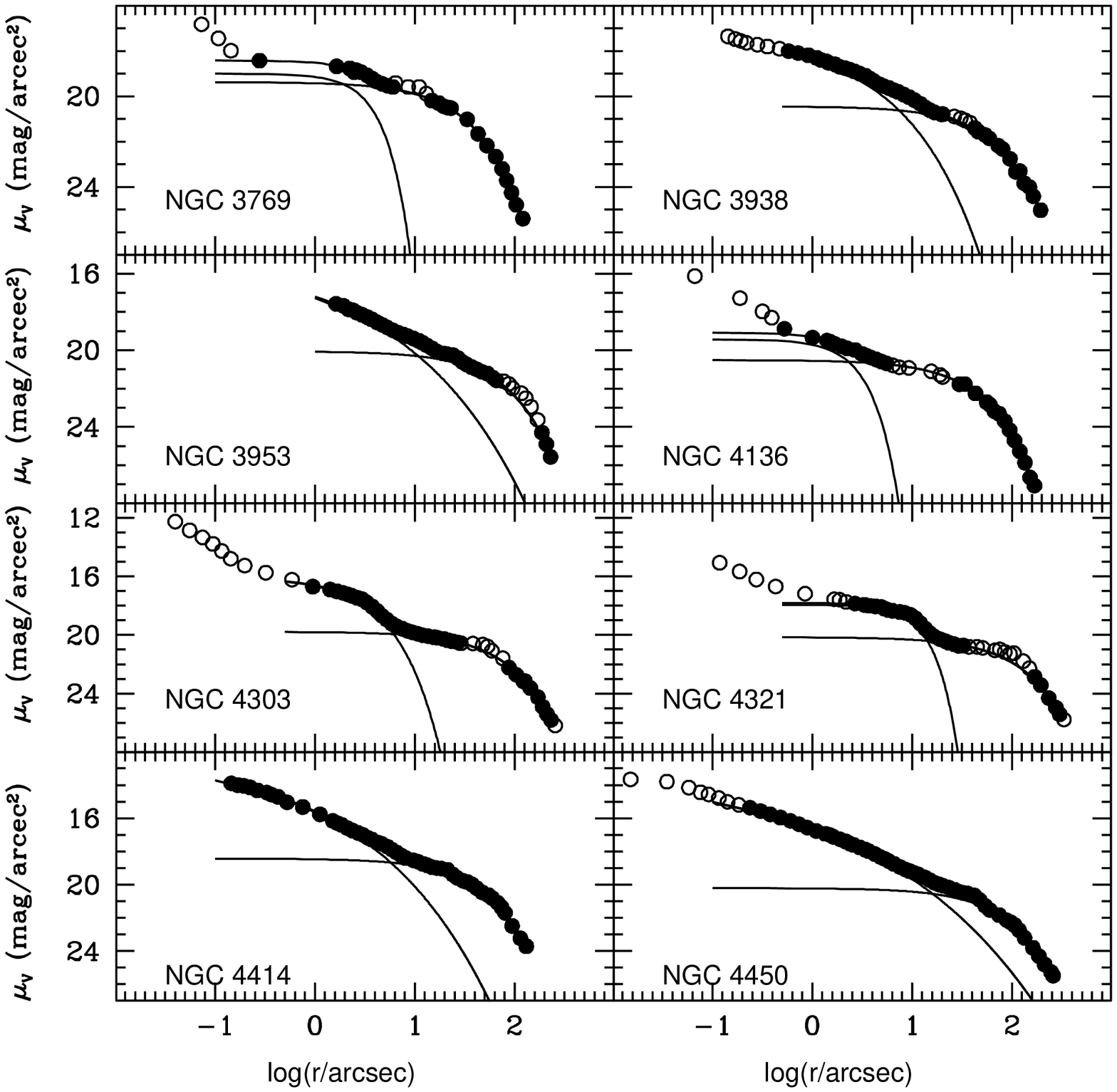}
\caption{Above we show new V-band S\'ersic fits in this paper. Open symbols represent surface brightness isophotes; the filled symbols indicated data elements included in the S\'ersic decomposition. The black lines indicate the fitted function for each galaxy.\label{fig:newprofs}}
\end{figure*}

In two galaxies (M~101 and IC~342) there was not sufficient coverage
in the optical two constrain a bulge-disk decomposition. We therefore
adduce 2MASS $J$-band data to extend the dynamic fitting range. We
stress that we only use the fits in this paper for bulge S\'ersic
index, and furthermore this S\'ersic index is only used two help
pseudobulge classification. In both of these galaxies the nuclear
morhology and IR activity strongly indicate that the bulge is a
pseudobulge, and the fitted S\'ersic index does not conflict with this
result.

We carry out a bulge-disk decomposition on each galaxy in our sample
by fitting the following equation (Eq.~\ref{eq:sersic}) to the major
axis surface brightness profiles by method of least-squares,
\begin{equation}
I(r)=I_e \exp\left[-b_n\left ( \left( \frac{r}{r_e}\right )^{1/n_b}-1 \right )\right] + I_d \exp\left [ \frac{r}{h}\right] \label{eq:sersic}
\end{equation}
where $b_n$ is a constant function of $n$ given in many publications \cite[e.g.][]{ciotti1999},
\begin{equation}
b_n\approx 2n - \frac{1}{3} + \frac{4}{405n} + \frac{46}{25515n^2}+\frac{131}{1148175n^3}+O(n^{-4}), \label{eq:bn}
\end{equation}
and the surface brightness of the bulge and disk are converted to
magnitudes respectively as follows $mu_e=-2.5\log(I_e)$ and
$mu_d=-2.5\log(I_d)$.

The decomposition is carried out on a major axis profile using the
mean isophote brightness. It does not take ellipticity into account
during the fitting. Thus, we take the mean ellipticity for each
component and adjust the luminosity accordingly:
$L=(1-\bar{\epsilon})L_{\mathrm{fit}}$. The radius of the component is
defined as the radius range within which that component dominates the
light of the profile.

Bars, rings, lenses, and similar features do not conform to the smooth
nature of Eq.~1, hence we carefully exclude regions of the profile
perturbed by such structures from the fit. This is a risky procedure,
as it requires selectively removing data from a galaxy's profile, and
undoubtedly has an effect on the resulting parameters. For those
galaxies in which a bar is present, it is our assumption that removing
the bar from the fit provides the best estimation of the properties of
the underlying bulge and disk. If a region is not included in a fit we
show that in the figure by using open symbols. This procedure is
described extensively in \cite{fisherdrory2008}.

\setcounter{figure}{12}
\begin{figure*}
\includegraphics[width=\textwidth]{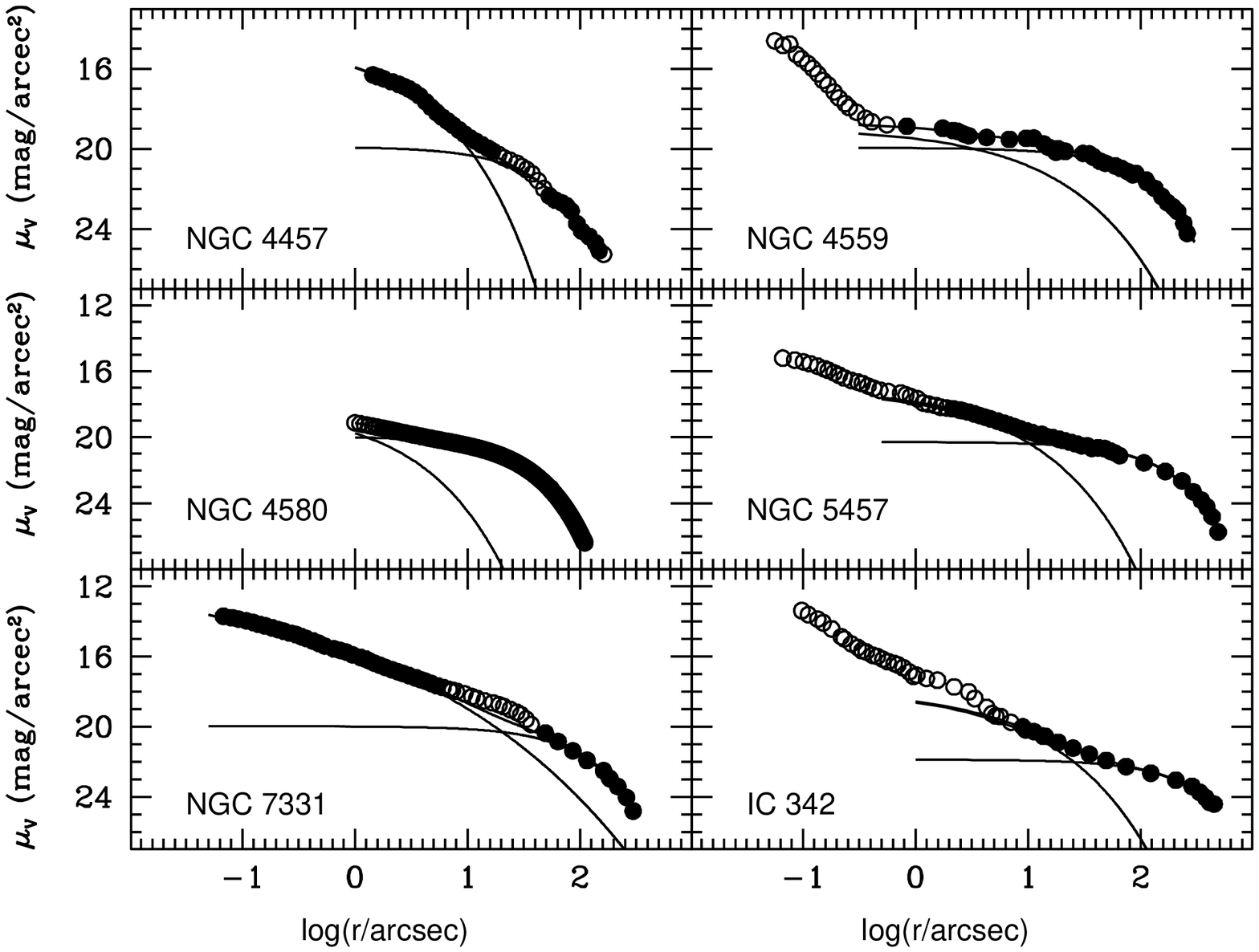}
\caption{Above we show new V-band S\'ersic fits in this paper. Open symbols represent surface brightness isophotes; the filled symbols indicated data elements included in the S\'ersic decomposition. The black lines indicate the fitted function for each galaxy.\label{fig:newprofs}}
\end{figure*}

\begin{deluxetable}{lcccccc}
  \tablewidth{0pt} \tablecaption{Parameters Of New Decompositions}
  \tablehead{\colhead{Identifier} & \colhead{$n_b$} & \colhead{$\mu_e$\tablenotemark{a}} &
    \colhead{$r_e$}
    & \colhead{$\mu_d$\tablenotemark{a}} &\colhead{h} & \colhead{Data Source\tablenotemark{a}} \\
    \colhead{} & \colhead{} & \colhead{(V-mag arcsec$^{-2}$)} &
    \colhead{arcsec} & \colhead{(V-mag arcsec$^{-2}$)}
    &\colhead{(arcsec)} & \colhead{} } \startdata	
NGC0925 & 1.2 $\pm$ 0.2 & 24.3 $\pm$ 0.3 & 23.8 $\pm$ 3.5 & 23.6 $\pm$ 0.3 & 101.9 $\pm$ 9.5& 1,2 \\
NGC1433 & 0.9 $\pm$ 0.1 & 17.8 $\pm$ 0.4 & 5.1 $\pm$ 1.0 & 19.8 $\pm$ 0.2 & 48.2 $\pm$ 1.8 & 1,4  \\
NGC1617 & 2.0 $\pm$ 0.2 & 18.8 $\pm$ 0.6 & 5.3 $\pm$ 3.2 & 19.4 $\pm$ 0.1 & 33.0 $\pm$ 0.6 & 1,4 \\
NGC1672 & 1.2 $\pm$ 0.1 & 17.7 $\pm$ 0.3 & 4.9 $\pm$ 0.8 & 19.5 $\pm$ 0.2 & 36.4 $\pm$ 0.8 & 1,5 \\
NGC2403 & 1.5 $\pm$ 0.6 & 22.6 $\pm$ 1.0 & 48.3 $\pm$ 34.9 & 20.2 $\pm$ 0.6 & 127.7 $\pm$ 14.0 & 1,6 \\
NGC3184 & 1.8 $\pm$ 0.5 & 20.1 $\pm$ 1.1 & 4.3 $\pm$ 1.2 & 20.5 $\pm$ 0.3 & 71.2 $\pm$ 6.1 &  1,2,3,7\\
NGC3675 & 3.2 $\pm$ 0.6 & 21.1 $\pm$ 0.7 & 30.9 $\pm$ 21.1 & 19.7 $\pm$ 0.4 & 43.4 $\pm$ 4.0 & 1,3\\
NGC3726 & 1.9 $\pm$ 0.3 & 20.8 $\pm$ 0.8 & 7.4 $\pm$ 5.6 & 20.4 $\pm$ 0.2 & 46.2 $\pm$ 1.0 & 3, 8, 13\\
NGC3769 & 0.5 $\pm$ 0.1 & 19.8 $\pm$ 0.2 & 2.6 $\pm$ 0.2 & 19.4 $\pm$ 0.1 & 20.9 $\pm$ 0.2 & 1, 3\\
NGC3938 & 1.7 $\pm$ 0.3 & 20.8 $\pm$ 0.7 & 8.0 $\pm$ 2.7 & 20.4 $\pm$ 0.2 & 44.7 $\pm$ 1.3 & 1,2,3\\
NGC3953 & 2.7 $\pm$ 0.4 & 20.7 $\pm$ 1.1 & 13.0 $\pm$ 10.4 & 20.1 $\pm$ 0.2 & 46.3 $\pm$ 0.9 & 3,8\\
NGC4136 & 0.6 $\pm$ 0.6 & 20.3 $\pm$ 1.7 & 2.0 $\pm$ 0.6 & 20.5 $\pm$ 0.2 & 27.6 $\pm$ 0.7 & 1,3,8\\
NGC4303 & 1.0 $\pm$ 0.1 & 17.8 $\pm$ 0.5 & 2.8 $\pm$ 0.7 & 19.8 $\pm$ 0.1 & 40.7 $\pm$ 0.7 & 1,3,8\\
NGC4321 & 0.5 $\pm$ 0.1 & 18.7 $\pm$ 0.2 & 7.7 $\pm$ 0.5 & 20.2 $\pm$ 0.1 & 62.7 $\pm$ 1.3 & 1,3\\
NGC4414 & 2.7 $\pm$ 0.4 & 17.8 $\pm$ 0.9 & 3.9 $\pm$ 1.4 & 18.4 $\pm$ 0.2 & 26.4 $\pm$ 0.8 & 1,2,3\\
NGC4450 & 3.7 $\pm$ 0.3 & 20.8 $\pm$ 0.8 & 18.1 $\pm$ 9.2 & 20.2 $\pm$ 0.3 & 50.4 $\pm$ 1.9 & 1,2,3\\
NGC4457 & 1.7 $\pm$ 0.5 & 17.9 $\pm$ 1.7 & 4.5 $\pm$ 2.1 & 19.9 $\pm$ 0.8 & 26.7 $\pm$ 3.6 & 1,3,10,11\\
NGC4559 & 1.9 $\pm$ 0.8 & 22.6 $\pm$ 1.4 & 33.8 $\pm$ 28.8 & 19.9 $\pm$ 0.3 & 69.4 $\pm$ 3.4 & 1,3\\
NGC4580 & 1.6 $\pm$ 0.6 & 21.4 $\pm$ 2.3 & 3.3 $\pm$ 6.0 & 19.9 $\pm$ 0.2 & 18.5 $\pm$ 0.4 &  3\\
NGC5457 & 1.8 $\pm$ 0.5 & 20.7 $\pm$ 1.2 & 11.9 $\pm$ 5.5 & 20.3 $\pm$ 0.2 & 102.5 $\pm$ 2.6 & 1,3,7,12 \\
NGC7331 & 4.5 $\pm$ 0.5 & 20.7 $\pm$ 0.9 & 24.4 $\pm$ 9.5 & 20.0 $\pm$ 0.5 & 66.3 $\pm$ 4.8 & 1,3\\
IC0342 & 1.9 $\pm$ 0.4 & 21.6 $\pm$ 1.3 & 21.1 $\pm$ 22.0 & 21.9 $\pm$ 0.2 & 189.7 $\pm$ 9.3 & 1,12\\

 \enddata
 \tablenotetext{a}{We are only interested in S\'ersic index. Thus
   magnitudes have not been corrected for Galactic extinction.}
 \tablenotetext{b}{Data Source References are as follows: 1--HST
   Archive; 2--\cite{sings}; 3--\cite{sdssdr7}; 4--\cite{hameed1999};
   5--\cite{kuchinski2000}; 6--\cite{larsen1999};
   7--\cite{knapen2004}; 8--\cite{frei1996}; 9--\cite{cheng1997};
   10--\cite{koopmann2001}; 11--\cite{eskridge2000}; 12 --
   \cite{2mass}; 13 --\cite{tully1996}. }
 \end{deluxetable}

\end{document}